\documentclass[twocolumn]{aastex63}
\usepackage{amsmath}
\usepackage{gensymb}

\usepackage{xcolor}
\usepackage{verbatim}
\definecolor{medium-blue}{rgb}{0,0,1}

\definecolor{my_color}{HTML}{3a18b1}
\definecolor{new_color}{HTML}{CF0000}
\definecolor{new_black}{HTML}{000000}
\newcommand\bmaroon[0]{\textcolor{new_black}}

\newcommand\ms{\ensuremath{\text{m}\,\text{s}^{-1}}}
\newcommand\masyr{\ensuremath{\text{mas}\,\text{yr}^{-1}}}

\received{December 9, 2020}
\revised{March 26, 2021}
\accepted{April 15, 2021}
\shortauthors{Venner et al.}

\begin{document}

\title{True masses of the long-period companions to HD 92987 and HD 221420 from Hipparcos--Gaia astrometry}

\correspondingauthor{Alexander Venner}
\email{AlexanderVenner@gmail.com}

\author[0000-0002-8400-1646]{Alexander Venner}
\affiliation{Monmouth, UK}

\author[0000-0001-7246-5438]{Andrew Vanderburg}
\affiliation{Department of Astronomy, The University of Wisconsin-Madison, Madison, WI 53706, USA}

\author[0000-0003-3904-7378]{Logan A. Pearce}
\affiliation{Steward Observatory, University of Arizona, Tucson, AZ 85721, USA}



\begin{abstract}

The extensive timespan of modern radial velocity surveys has made the discovery of long-period substellar companions more common in recent years, however measuring the true masses of these objects remains challenging. Astrometry from the Gaia mission is expected to provide mass measurements for many of these long-period companions, but this data is not yet available. However, combining proper motion data from Gaia DR2 and the earlier Hipparcos mission makes it possible to measure true masses of substellar companions in favourable cases. In this work, we combine radial velocities with Hipparcos-Gaia astrometry to measure the true masses of two recently discovered long-period substellar companion candidates, HD~92987~B and HD~221420~b. \bmaroon{In both cases,} we find that the true masses are significantly higher than \bmaroon{implied} by radial velocities alone\bmaroon{. A $2087 \pm 19$ \ms{}} astrometric signal reveals that HD~92987~B is not close to its 17 $M_J$ minimum mass but is instead a \bmaroon{$0.2562 \pm 0.0045$} $M_\odot$ star viewed at a near-polar \bmaroon{orbital} inclination, whereas the $22.9 \pm 2.2$ $M_J$ HD~221420~b can be plausibly interpreted as a high-mass ``super-planet" or a low-mass brown dwarf. With semi-major axes of $\sim$10 AU both companions are interesting targets for direct imaging, and HD~221420~b in particular would be a benchmark metal-rich substellar object if it proves possible to directly detect. Our results demonstrate the power of Hipparcos-Gaia astrometry for studying long-period planet and brown dwarf candidates discovered from radial velocity surveys.

\end{abstract}



\section{Introduction} \label{sec:intro}

The two and a half decades since the discovery of 51~Pegasi~b by \citet{51peg} have seen a continuous and rapid expansion in our knowledge of planets beyond the solar system. Although much focus was initially placed on the unexpected existence of 51 Peg b-like ``hot Jupiters", over time it has become clear that exoplanets are both diverse and abundant (\citealt{Winn15}, \citealt{Perryman18}, \citealt{Hsu20}).

The long time baselines of radial velocity (RV) observations now available for many stars allows for the discovery of exoplanets with orbital periods comparable to those of the giant planets found in the Solar System. Previous studies indicate that these planets are more common than giant planets on shorter orbital periods, with giant planets beyond $P \gtrsim 300$ days significantly outnumbering those with $P \lesssim 300$ days \citep{Wittenmyer20}, although there is evidence that this trend is non-linear such that giant planets become less frequent at much longer periods \citep{Fernandes19}.

Although radial velocities are the source of a large portion of the knowledge of planets between 1 - 20 AU, the method has its limitations. The requirement that a significant part of the orbit must be covered to constrain orbital parameters means that a star must be observed for decades in order to discover long-period giant planets, and the well-known $\sin i$ degeneracy leaves the true mass of companions unknown unless it can be inferred by other means. Such complementary data is scarce; while direct imaging is capable of exploring well into the planetary regime for young stars (e.g. \citealt{51Erib}, \citealt{Bowler16}), imaging instruments are not yet capable of detecting planets orbiting older stars like many of those discovered using radial velocities. While astrometry can be used to constrain orbital inclinations, currently available astrometric data is generally insufficiently precise to detect planets; the main exception is astrometry from the Hubble Space Telescope, which has been used to constrain the true masses of a small selection of radial velocity planets (see \citealt{Benedict17} for a review). Astrometric data from the Gaia mission \citep{Gaia} is expected to greatly improve on this situation with a projected yield of thousands of new astrometric planet detections \citep{Perryman14}, but this data will not be made available until the release of the Gaia DR4 subsequent to completion of the nominal Gaia mission. As a result, the true masses of many companions discovered using radial velocities remain unknown.

The combination of proper motion data from Gaia and Hipparcos \citep{Hipparcos} offers a workaround for these limitations. The proper motion of an astrometric source effectively circumscribes the relative tangential velocity of the object over the span of observations. Proper motions of $\sim$1.3 billion stars are available in the Gaia DR2 \citep{GaiaDR2}, and these can then be cross-matched with the proper motions of the $\sim$118000 stars in the Hipparcos catalogue to produce two measurements of tangential velocity separated by $\sim$24 years. When combined with other data (e.g. radial velocities, direct imaging) these proper motions can be used to constrain the parameters of orbiting companions, including inclinations. Past publications which applied this technique to exoplanets include \citet{Snellen18}, \citet{Dupuy19}, and \citet{Nielsen20} for $\beta$~Pictoris~b, \citet{Feng19} for Epsilon~Indi~b, \citet{DeRosa20.51Eri} for 51~Eridani~b, \citet{DeRosa19} for a suspected planetary companion to TW~Piscis~Austrini, \citet{Xuan20} for $\pi$~Mensae~b and HAT-P-11~c, \citet{Damasso20} and \citet{DeRosa20.piMen} for $\pi$~Mensae~b alone, and finally \citet{Kervella20} for Proxima~Centauri~c. The technique has also been successfully applied to stars with brown dwarf and stellar companions (\citealt{Calissendorff18, Brandt19, Brandt19.gl229, Grandjean19, DeRosa19.HR1645, Bonavita20, Maire20.HD72946, Maire20.HD19467}), which often produce stronger signals owing to their higher masses.

In this work, we use Hipparcos-Gaia astrometry to measure the inclinations and true masses of two recently discovered long-period substellar companion candidates, HD~92987~B and HD~221420~b.

\section{Targets} \label{sec:targets}

\begin{deluxetable*}{ccc}[t]
\centering
\tablecaption{Observed and inferred parameters of the target stars \bmaroon{adopted} in this study.\label{table:starparams}}
\tablehead{\colhead{Parameter} & \colhead{HD 92987} & \colhead{HD 221420}}
\startdata
Spectral Type & G2/3IV-V $^{(1)(a)}$ & G2IV-V $^{(2)}$ \\ 
$V$ [mag] & $7.02 \pm 0.01$ $^{(3)}$ & $5.806 \pm 0.009$ $^{(3)}$ \\ 
Parallax $\varpi$ [mas] & $22.942 \pm 0.035$ $^{(4)}$ & $32.083 \pm 0.039$ $^{(4)}$ \\ 
Luminosity $L_*$ [$L_\odot$] & $2.546 \pm 0.006$ $^{(4)}$ & $4.008 \pm 0.008$ $^{(4)}$ \\ 
Temperature $T_{eff}$ [K] & $5774 \pm 44$ $^{(5)}$ & $5847 \pm 22$ $^{(6)}$ \\ 
$\lbrack$Fe/H$\rbrack$ [dex] & $0.05 \pm 0.03$ $^{(5)}$ & $0.33 \pm 0.02$ $^{(6)}$ \\ 
log $g$ [cgs] & $4.06 \pm 0.06$ $^{(5)}$ & $4.03 \pm 0.03$ $^{(6)}$ \\ 
\hline 
Mass $M_*$ [$M_\odot$] & \bmaroon{$1.043 \pm 0.012 \text{ (stat.)} \pm 0.013 \text{ (sys.)}$} $^{(7)}$ & \bmaroon{$1.351 \pm 0.012 \text{ (stat.)} \pm 0.005 \text{ (sys.)}$} $^{(7)}$ \\ 
Radius $R_*$ [$R_\odot$] & $1.569 \pm 0.012$ $^{(7)}$ & $1.947 \pm 0.013$ $^{(7)}$ \\ 
Age [Gyr] & \bmaroon{$7.98 \pm 0.25 \text{ (stat.)} \pm 0.65 \text{ (sys.)}$} $^{(7)}$ & \bmaroon{$3.65 \pm 0.13 \text{ (stat.)} \pm 0.19 \text{ (sys.)}$} $^{(7)}$ \\ 
\enddata
\tablerefs{(1) \citet{Houk82}; (2) \citet{Gray06}; (3) \citet{Tycho2}; (4) \citet{GaiaDR2}; (5) \citet{Valenti05}; (6) \citet{Sousa08}; (7) This work.}
\tablecomments{(a) The luminosity classification differs from the class V of \citet{Houk82} as we interpret HD~92987 as a turnoff star (see text).}
\end{deluxetable*}

The adopted parameters of the two target stars are listed in Table \ref{table:starparams}. Observable parameters were taken from the literature, while the masses, radii, and ages were calculated using the PARAM 1.5 online interface\footnote{\url{http://stev.oapd.inaf.it/cgi-bin/param}} (\citealt{daSilva06}; \citealt{Rodrigues14}, \citeyear{Rodrigues17}) using the ``1-step" method and the PARSEC isochrones \citep{Bressan12}, with the physical parameters in Table \ref{table:starparams} and the stars' BVJHK magnitudes as input priors. \bmaroon{For the masses and ages we add in estimates for systematic uncertainties generated using \texttt{kiauhoku} \citep{Tayar20}.} More detailed discussion of the target stars is given below.

\subsection{HD 92987} \label{subsec:targets:HD92987}

HD~92987 is a near-solar metallicity G2/3 type star located at a distance of 43.59 $\pm$ 0.07 parsecs from the Solar System. Although previously designated as luminosity class V by \citet{Houk82}, based on the large radius and luminosity in Table \ref{table:starparams} we infer that the star is at the main sequence turnoff, and thus adopt a luminosity class of IV-V for the star in this study. HD~92987 is not known to have any widely separated stellar companions.

A long-period substellar companion candidate orbiting HD~92987 was independently discovered by \citet{Rickman19} and \citet{Kane19} using radial velocities from the CORALIE (ESO) and UCLES (AAT) spectrographs, respectively. Although based on separate datasets, the two studies recovered concordant orbital parameters for the companion, with an orbital period of $\sim$10500 days ($\sim$29 years), a RV semi-amplitude of $\sim$150 \ms{}, and a minimum mass of $\sim$17 M$_J$. This mass is above the canonical deuterium burning limit of 13 M$_J$ \citep{Boss03, Boss07} which motivated \citet{Rickman19} to refer to the companion as a brown dwarf candidate, whereas \citet{Kane19} refer to the companion as a planet.

HD~92987 also notably appears in \citet{Frankowski07} who searched for binary stars by comparing the short term proper motions from Hipparcos \citep{Hipparcos} and the long-term proper motions from Tycho-2 \citep{Tycho2}. A star showing significant differences between the two proper motions may be perturbed by a companion, and the stars that display this variability in proper motion were named ``$\Delta\mu$ binaries" by \citet{Wielen99}. For HD~92987, \citet{Frankowski07} measured ${\chi}^2=22.55$ and $P({\chi}^2)=0.99999$ for the hypothesis of constant proper motion, exceeding their adopted confidence level of $P({\chi}^2)=0.9999$, and they thus identified the star as a candidate proper motion binary. If this can be associated with the companion detected in the RV data then it suggests that the astrometric signal produced by HD~92987~B is relatively large, greater than expected from the $\sim$17 M$_J$ minimum mass of the companion.

\subsection{HD 221420}

HD~221420 is a metal-rich, naked-eye G3 star positioned around the main sequence turnoff lying at a distance of 31.17 $\pm$ 0.04 parsecs. The star has a relatively small proper motion, and based on astrometry from Gaia DR2 \citet{BailerJones18} found that it made its closest approach to the Solar System $1.137 \pm 0.011$ million years ago at a distance of $2.874 \pm 0.037$ parsecs, although the proper motions used in that work are affected by orbital motion so the reported values may be slightly inaccurate. HD~221420 was not previously known to have any stellar companions \citep{Eggleton08}, but in this work we identify a codistant M-dwarf that may be bound (see Appendix \ref{appendix:A}).

A long-period planet candidate orbiting the star was discovered by \citet{Kane19} using AAT radial velocities spanning nearly 18 years. Those authors found an orbital period of $22482^{+4200}_{-4100}$ days ($61.55^{+11.50}_{-11.23}$ years), a RV semi-amplitude of $54.7^{+4.2}_{-3.6}$ \ms{}, and a minimum mass of $9.7^{+1.1}_{-1.0}$ M$_J$ for the companion, although the available data covers only one RV extremum so the uncertainties on the companion parameters are large.

\section{Method} \label{sec:method}

The methodology employed in this work is chiefly inspired by those developed by \citet{Brandt19} and \citet{Feng19}; our method was developed independently of that of \citet{Xuan20} but shares its basic features. \bmaroon{We provide a pedagogical description of our methodology here to increase the reproducibility of our results and} so that Hipparcos-Gaia astrometry can be more widely used in future works.

\subsection{Data} \label{subsec:data}

For the radial velocity data, we incorporate the previously published CORALIE and AAT RVs from \citet{Rickman19} and \citet{Kane19} for HD~92987, and the AAT RVs from the latter publication  for HD~221420. In the case of HD~92987, which was discovered independently by the two groups, this is the first time that the two sets of radial velocity data have been analysed together.

As well as being observed by the AAT, HD~221420 is also a target of the HARPS (ESO) spectrograph, specifically on the GTO sample according to its inclusion in \citet{Sousa08}. HARPS RVs of this star have not previously been used in the literature, but this data is publicly available for access using the ESO science archive.%
\footnote{\url{http://archive.eso.org/wdb/wdb/adp/phase3_spectral/form?phase3_collection=HARPS}} For the measurements taken prior to the 2015 HARPS upgrade we make use of the re-reduction performed by \citet{Trifonov20}; however, the ESO archive contains some \bmaroon{observations from 2018} that were not included in that work, so we instead extract the post-upgrade RVs from the archive using the standard DRS. Additionally, we take nightly bins of all of the HARPS data. The HARPS radial velocities used in this work are provided in Appendix \ref{appendix:B}.

The proper motion data used in this work is derived from the Hipparcos-Gaia Catalog of Accelerations (HGCA; \citealt{Brandt18}, with corrections provided by \citealt{Brandt18erratum}). Although we acknowledge the similar work of \citet{Kervella19}, we choose the HGCA here as it includes a localised cross-calibration, uses a linear combination of the two Hipparcos reductions (\citealt{Hipparcos, HipparcosNew}) which was demonstrated to be superior to using either reduction alone by \citet{Brandt18}, and additionally provides precise epochs for the astrometric observations which are required by the model used here (see Section \ref{subsec:interp}). Still, despite these differences the proper motion data for the two targets studied here does not differ greatly between the two works, so the choice of source does not significantly affect our results.

As described in \citet{Brandt19}, the astrometric data consists of three effectively independent sets of proper motion measurements:

\begin{itemize}
	\item The Hipparcos proper motion, $\mu_{H}$, measured near to epoch 1991.25;
	\item The Gaia DR2 proper motion, $\mu_{G}$, measured near to epoch 2015.5;
	\item The Hipparcos-Gaia mean proper motion, $\mu_{HG}$, calculated from the difference between the sky positions observed by the two telescopes $\sim$24 years apart.
\end{itemize}

This last measurement is referred to as the ``scaled positional difference" in \citet{Brandt19} and the ``mean motion vector" in \citet{Xuan20}; we simply refer to it as the ``Hip-Gaia proper motion" in this work. Hip-Gaia proper motion measurements are often extraordinarily precise, and when combined with the Gaia DR2 proper motions, $\Delta\mu$ anomalies can often be confidently detected down to a level of $\sim$0.1-0.2 \masyr{} ($\sim$15-30 \ms{} at a distance of 30 pc).

\subsection{Model: Radial Velocities}

For a two-body system, the velocity of the primary along the line of sight over time can be expressed as

\begin{equation}
v_{RV}(t)=K[\cos(\nu(t)+\omega)+e\cos\omega]
\:,
\end{equation}

where $\nu(t)$ is the true anomaly at time $t$, $e$ is the orbital eccentricity, $\omega$ is the argument of periastron for the primary's orbit, and $K$ is the radial velocity semi-amplitude given by

\begin{equation}
K=\frac{2\pi}{P}\frac{a_*\sin i}{(1-e^2)^{1/2}}
\:,
\end{equation}

where $P$ is the orbital period, $a_*$ is the semi-major axis of the primary around the barycentre, and $i$ is the orbital inclination (\citealt{Perryman18} Equations 2.22, 2.23). $K$ can also be expressed in a form involving the relative, primary-to-secondary semi-major axis $a$:

\begin{equation}
\label{equation:K}
K=\sqrt{\frac{G}{(m_A+m_B)a(1-e^2)}}m_B\sin i
\:,
\end{equation}

where $G$ is the gravitational constant and $m_A$ and $m_B$ are the masses of the primary and secondary respectively (\citealt{Lovis10} Equation 12).

\subsection{Model: Astrometry}

\subsubsection[Direct part]{Directly observed astrometry (Hipparcos, Gaia)} \label{subsec:direct}

The proper motion of a star with a companion can be approximately described with two components, a constant barycentric term ($\mu_{bary}$) and a variable orbital term ($\Delta\mu$). We focus on the latter component in this section.

The velocity of the primary of a two-body system in the orbital plane can be defined as

\begin{equation}
\label{equation:vxy}
\begin{bmatrix}
v_x(t) \\
v_y(t)
\end{bmatrix}
=\sqrt{\frac{Gm_B^2}{(m_A+m_B)a(1-e^2)}}
\begin{bmatrix}
-\sin\nu(t) \\
\cos\nu(t)+e
\end{bmatrix}
\end{equation}

(\citealt{Feng19} Equation 5, corrected\footnote{The equation given in \citet{Feng19} is missing the square root (F. Feng, personal communication).}). The left-hand side of this equation can be easily related to the $K$ parameter of Equation \ref{equation:K} by

\begin{equation}
\label{equation:kappa}
\sqrt{\frac{Gm_B^2}{(m_A+m_B)a(1-e^2)}} = \frac{K}{\sin i}
\:.
\end{equation}

We refer to this parameter as $\kappa$ in this work, such that Equation \ref{equation:vxy} is equivalent to

\begin{equation}
\begin{bmatrix}
v_x(t) \\
v_y(t)
\end{bmatrix}
=\kappa
\begin{bmatrix}
-\sin\nu(t) \\
\cos\nu(t)+e
\end{bmatrix}
\:,
\end{equation}

$\kappa$ thus being equivalent to the $A$ of \citet{Xuan20}. To convert these equations into tangential velocities comparable to the observations they must be rotated into the observer's reference frame, which follows

\begin{equation}
\label{equation:vRA}
\begin{split}
v_{RA}(t)=-&v_x(t)[\cos\omega\sin\Omega+\sin\omega\cos\Omega\cos i]\:- \\
&v_y(t)[-\sin\omega\sin\Omega+\cos\omega\cos\Omega\cos i]
\:,
\end{split}
\end{equation}

\begin{equation}
\label{equation:vDec}
\begin{split}
v_{Dec}(t)=-&v_x(t)[\cos\omega\cos\Omega-\sin\omega\sin\Omega\cos i]\:- \\
&v_y(t)[-\sin\omega\cos\Omega-\cos\omega\sin\Omega\cos i]
\:,
\end{split}
\end{equation}

where $\Omega$ is the longitude of node for the \textit{secondary} orbit. While it may be intuitively preferable to use the longitude of node for the orbit of the primary for consistency with the argument of periastron, we follow the convention set by direct imaging by using the node of the secondary, which produces the negative signs in the above equations. Finally, the unit conversion between \ms{} and \masyr{} is

\begin{equation}
\label{equation:muRA}
\Delta\mu_{RA} \simeq \frac{\varpi}{4740.5} v_{RA}
\:,
\end{equation}

\begin{equation}
\label{equation:muDec}
\Delta\mu_{Dec} \simeq \frac{\varpi}{4740.5} v_{Dec}
\:,
\end{equation}

where $\mu$ is the proper motion in \masyr{}, $\varpi$ is the parallax in mas, and the term in the denominator is derived from

\begin{equation}
\frac{\text{AU}}{\text{m}}\frac{\text{s}}{\text{yr}} \simeq 4740.5
\:,
\end{equation}

i.e. the number of metres in 1 AU multiplied by the number of years in 1 second (equivalent to division by the number of seconds in 1 year).

While the preceding equations would be sufficient if the Hipparcos and Gaia proper motions were instantaneous measurements, in practice the values are averages for multiple years of astrometric observations - $\sim3.36$ years for Hipparcos \citep{Hipparcos} and $\sim1.75$ years for Gaia DR2 \citep{GaiaDR2astrometry}. As a result, it is necessary to resample the model proper motions over the observational epochs for both instruments. The Hipparcos epochs can be derived in a straightforward manner from the Hipparcos Intermediate Astrometric Data or the Hipparcos Epoch Photometry Annex \citep{HipparcosPhot}, but the Gaia epochs are not available in the DR2. Instead, as in previous works (e.g. \citealt{Calissendorff18, Nielsen20, Xuan20}), we use the Gaia Observation Forecast Tool\footnote{\url{https://gaia.esac.esa.int/gost/}} to obtain estimates for the Gaia observation times, which should be sufficiently accurate approximations of the true epochs for the purposes of resampling. As in \citet{Xuan20}, we remove epochs within 1 day of each other to avoid times with inflated statistical weights. We then calculate model $\Delta\mu$ values for each individual epoch, then take the average for each instrument to find our model values for $\Delta\mu_H$ and $\Delta\mu_G$.

\subsubsection[Interpolated part]{Time-averaged astrometry (Hip-Gaia)} \label{subsec:interp}

Unlike \bmaroon{some} previous studies that utilised Hipparcos-Gaia astrometry (e.g. \citealt{Brandt19}\bmaroon{, although see \citealt{Brandt19.gl229, Currie20} for counterexamples}), we do not normalise the proper motion data to the Hip-Gaia value. This allows - or rather requires - the components of the barycentric proper motion to be incorporated as variable parameters, which increases the computational cost but provides physically meaningful parameter constraints that can be used elsewhere (for example, the barycentric proper motions of HD~221420 are used in Appendix \ref{appendix:A} for comparison with a suspected companion). Doing this requires direct modelling of the Hip-Gaia proper motion, and as this measurement is averaged over $\sim$24 years a different approach is required than for modelling the Hipparcos and Gaia proper motions alone.

The Hip-Gaia proper motion is effectively the average proper motion of the star between two epochs, the Hipparcos observation time and the Gaia observation time, which includes a constant barycentric component and a variable orbital component. Unless the variation in proper motions is approximately linear between the two epochs it is impractical to use the expressions given in Section \ref{subsec:direct} to model the average proper motion; instead it is preferable to integrate over the time interval, as the orbital part of the Hip-Gaia proper motion can be expressed as

\begin{equation}
\label{equation:muHGint}
\Delta\bar\mu=\Delta\mu_{HG}=\frac{1}{t_G-t_H} \int_{t_H}^{t_G} \mu(t)
\:,
\end{equation}

where $t$ is the epoch of an observation, and $H$ and $G$ refer to Hipparcos and Gaia respectively. As proper motions express tangential velocities, the integral \bmaroon{in Equation \ref{equation:muHGint}} is equivalent to the difference between the orbital positions of the star at the two epochs\bmaroon{, resulting in the following expression}:

\begin{equation}
\label{equation:muHG}
\Delta\mu_{HG}=\frac{r_G-r_H}{t_G-t_H}
\:,
\end{equation}

where $r$ is the position of the star relative to the barycentre. Thus, equations in position rather than velocity are required here.

In the orbital plane, the position of the primary of a two-body system with respect to the barycentre is defined as

\begin{equation}
r(t)=
\begin{bmatrix}
x(t) \\
y(t) \\
\end{bmatrix}
=a_*
\begin{bmatrix}
\cos E(t)-e \\
\sqrt{1-e^2}\sin E(t)
\end{bmatrix}
\:,
\end{equation}

where $E(t)$ is the eccentric anomaly at time $t$ (\citealt{Feng19} Equation 4). $a_*$ is related to $a$ by

\begin{equation}
a_*=\frac{m_B}{m_A+m_B}a
\:,
\end{equation}

i.e. $a$ is the semi-major axis for $m_A+m_B$, and $a_*$ is the proportion of that value generated by $m_B$.

The method for rotation into the observer's reference frame is the same as in Equations \ref{equation:vRA} and \ref{equation:vDec}:

\begin{equation}
\begin{split}
r_{RA}(t)=-&x(t)[\cos\omega\sin\Omega+\sin\omega\cos\Omega\cos i]\:- \\
&y(t)[-\sin\omega\sin\Omega+\cos\omega\cos\Omega\cos i]
\:,
\end{split}
\end{equation}

\begin{equation}
\begin{split}
r_{Dec}(t)=-&x(t)[\cos\omega\cos\Omega-\sin\omega\sin\Omega\cos i]\:- \\
&y(t)[-\sin\omega\cos\Omega-\cos\omega\sin\Omega\cos i]
\:.
\end{split}
\end{equation}

These equations can then be combined with Equation \ref{equation:muHG} to produce the Hip-Gaia proper motions:

\begin{equation}
\Delta\mu_{HG,RA}=\frac{r_{RA}(t_{G,RA})-r_{RA}(t_{H,RA})}{t_{G,RA}-t_{H,RA}}
\:,
\end{equation}

\begin{equation}
\Delta\mu_{HG,Dec}=\frac{r_{Dec}(t_{G,Dec})-r_{Dec}(t_{H,Dec})}{t_{G,Dec}-t_{H,Dec}}
\:.
\end{equation}

The required epochs of astrometric observations are provided in the HGCA. As discussed in \citet{Brandt18} the effective astrometric epochs for a particular instrument are not identical between right ascension and declination, so a total of four epochs are required for each star.

For completeness we note that a unit conversion from \ms{} and \masyr{} as in Equations \ref{equation:muRA} and \ref{equation:muDec} may be necessary depending on the units used, but the model used here accounts for this by taking $a_*$ in units of mas.

\begin{deluxetable*}{lcccc}[ht!]
\centering
\tablecaption{Comparative parameters of $\pi$ Mensae b.\label{table:piMen}}
\tabletypesize{\scriptsize}
\tablehead{\colhead{Parameter} & \colhead{\citet{Xuan20}} & \colhead{\citet{Damasso20}} & \colhead{\citet{DeRosa20.piMen}} & \colhead{This work}}
\startdata
Period $P$ [days] & $2090.3 \pm 2.6$ & $2088.8 \pm 0.4$ & $2089.11^{+0.36}_{-0.37}$ & $2091.0 \pm 1.9$ \\
RV semi-amplitude $K$ [\ms{}] & -- & $196.1 \pm 0.7$ & $193.7 \pm 0.33$ & $192.8 \pm 1.4$ \\
Eccentricity $e$ & $0.644 \pm 0.003$ & $0.642 \pm 0.001$ & $0.6450 \pm 0.0011$ & $0.641 \pm 0.003$ \\
Argument of periastron (primary) $\omega_1$ [degrees] & $331.7 \pm 0.9$ $^{(a)}$ & $329.9 \pm 0.3$ & $331.15^{+0.24}_{-0.23}$ & $330.8 \pm 0.6$ \\
Time of periastron $T_p$ [JD] & $2456306.5 \pm 7.7$ & $2458388.6 \pm 2.2$ & $2452123.21^{+1.20}_{-1.18}$ & $2452122.58 \pm 1.65$ \\
Secondary minimum mass $m_2\sin i$ [$M_J$] & -- & $9.89 \pm 0.25$ & -- & $9.96 \pm 0.24$ \\
Relative semi-major axis $a$ [AU] & -- & $3.28 \pm 0.04$ & $3.308 \pm 0.039$ & $3.308 \pm 0.039$ \\
\hline
Orbital inclination $i$ [degrees] & $51.2^{+14.1}_{-9.8}$ & $45.8^{+1.4}_{-1.1}$ & $49.9^{+5.3}_{-4.5}$ & $50.8^{+12.5}_{-8.8}$ \\
Longitude of node (secondary) $\Omega_2$ [degrees] & $105.8^{+15.1}_{-14.3}$ & $108.8^{+0.6}_{-0.7}$ & $90.3^{+8.1}_{-8.0}$ $^{(b)}$ & $106.8^{+14.9}_{-15.0}$ \\
Secondary mass $m_2$ [$M_J$] & $12.9^{+2.3}_{-1.9}$ & $14.1^{+0.5}_{-0.4}$ & $13.01^{+1.03}_{-0.95}$ & $12.9^{+2.1}_{-1.7}$ \\
\hline
Barycentric RA proper motion $\mu_{bary,RA}$ [\masyr{}] & -- & -- & -- & $+310.52 \pm 0.02$ \\
Barycentric dec. proper motion $\mu_{bary,Dec}$ [\masyr{}] & -- & -- & -- & $+1049.62 \pm 0.03$ \\
\enddata
\tablecomments{(a) In \citet{Xuan20}, the radial (Z) direction is positive towards the observer, rather than the standard negative convention. In consequence their argument of periastron is rotated by 180$\degree$ relative to other results, so we give their value plus 180$\degree$ here. (b) The longitude of node in \citet{DeRosa20.piMen} is for the primary ($\Omega_1$), so we have applied a rotation of 180$\degree$ to their value.}
\end{deluxetable*}

\subsection{Model: Fitting} \label{subsec:modelfit}

In order to sample our \bmaroon{joint} RV-astrometry model, we use the Markov Chain Monte Carlo ensemble sampler \texttt{emcee~v3.0.2} \citep{emcee}. We use a total of $11+2N$ variable parameters where $N$ is the number of RV datasets per system, as follows:

\begin{itemize}
	\item Two externally constrained parameters, the parallax ($\varpi$) and the primary mass ($M_*$). We assume Gaussian priors on these parameters with the values listed in Table \ref{table:starparams}.
	\item Five parameters required to describe the radial velocity variations, namely the orbital period ($P$), the RV semi-amplitude ($K$), the orbital eccentricity and argument of periastron of the primary orbit ($e$, $\omega_1$) parametrised as $\sqrt{e}\sin\omega_1$ and $\sqrt{e}\cos\omega_1$ (see \citealt{EXOFAST} for discussion of different parametrisations of these parameters), and finally the time of periastron ($T_p$). In this work we refer to this grouping as the ``radial velocity parameters" as they are dominantly constrained by the RVs, but it should be noted that the proper motion data does provide ancillary constraints for these parameters, especially the orbital period.
	\item Four parameters to describe the proper motion data, being the orbital inclination ($i$), the longitude of node of the secondary orbit ($\Omega_2$), and the proper motions of the system barycentre in both co-ordinates ($\mu_{bary,RA}$, $\mu_{bary,Dec}$). We refer to this grouping as the ``astrometric parameters".
	\item Two parameters for each included radial velocity dataset, namely a constant offset ($\gamma$) which may be relative or barycentric depending on the dataset, and a jitter term ($\sigma_{\text{jit}}$) to account for excess noise in the RV data.
\end{itemize}

Other than the parallax and primary mass all parameters are sampled uniformly, and all are sampled linearly except the orbital period \bmaroon{and inclination, which are} sampled log-linearly \bmaroon{and according to $\sin i$ respectively}. As in \citet{Xuan20} the log-likelihood function of the model takes the form

\begin{equation}
\ln\mathcal{L}=-\frac{1}{2}(\chi^2_{RV}+\chi^2_\mu)
\:,
\end{equation}

where the radial velocity chi-squared is

\begin{equation}
\begin{split}
\chi^2_{RV}=
&\sum_{j=1}^{N_{dataset}}\sum_{k=1}^{N_{RV}}
\frac{(v_k-\mathcal{M}[v_k]-\gamma_j)^2}{\sigma^2_k+\sigma^2_{\text{jit},j}}\:+ \\
&\sum_{j=1}^{N_{dataset}}\sum_{k=1}^{N_{RV}}
\ln[2\pi(\sigma^2_k+\sigma^2_{\text{jit},j})]
\:,
\end{split}
\end{equation}

With $v_k$ and $\sigma_k$ being the $k^{\text{th}}$ radial velocity measurement and uncertainty in a dataset, $\mathcal{M}[v_k]$ being the corresponding model RV, and $\gamma_j$ and $\sigma_{\text{jit},j}$ being the model offset and jitter for the $j^{\text{th}}$ RV dataset (see \citealt{Howard14} Equation 1 and \citealt{Brandt19} Equation 16). The astrometric chi-squared is

\begin{equation}
\begin{split}
\chi^2_\mu=\sum_{j}^{N_{\mu}}\frac{1}{(1-\rho^2_j)}
\Bigg[
&\frac{(\mathcal{R}_{RA,j})^2}{\sigma^2_{RA,j}}+ \frac{(\mathcal{R}_{Dec,j})^2}{\sigma^2_{Dec,j}}\:- \\
&\frac{2\rho_j(\mathcal{R}_{RA,j})(\mathcal{R}_{Dec,j})}{\sigma_{RA,j}\times\sigma_{Dec,j}}
\Bigg]
\:,
\end{split}
\end{equation}

Where the sum is over the three proper motion measurements (i.e. Hipparcos, Gaia, and Hip-Gaia), $\rho_j$ is the right ascension-declination correlation coefficient for the $j^{\text{th}}$ measurement which is provided in the HGCA, and $\mathcal{R}$ stands for the O-C residuals such that

\begin{equation}
\mathcal{R}_{RA}=(\Delta\mu_{RA}-\mathcal{M}[\Delta\mu_{RA}]-\mu_{bary,RA})
\:,
\end{equation}

\begin{equation}
\mathcal{R}_{Dec}=(\Delta\mu_{Dec}-\mathcal{M}[\Delta\mu_{Dec}]-\mu_{bary,Dec})
\:.
\end{equation}

For the calculation of eccentric anomalies in our model we use the method of \citet{Zechmeister18}.

To initialise the model we take representative values for the RV parameters and hold them constant, then perform a least-squares regression on $\chi^2_\mu$ to determine initial values for the astrometric parameters. This set of parameters is then fed into the MCMC. \bmaroon{Convergence of the model was determined using the autocorrelation functions built into \texttt{emcee}; we considered convergence to have been reached once the chain was found to be 100 times longer than the estimated autocorrelation time averaged across all walkers for every variable parameter, a requirement which was typically reached within $1-2\times10^{5}$ steps depending on the system. To produce the posterior samples we discarded the first 25\% of the chain as burn-in and saved every hundredth step for each walker. We then extracted from the samples} the median and $68.3\%$ confidence intervals for the parameters used in the model as well as for physical parameters that can be derived from them such as the companion masses. The results of our models for the target systems are described in the proceeding section.

\section{Results} \label{sec:results}

\subsection[pi Mensae]{Comparative example: $\pi$ Mensae}

To demonstrate the efficacy of our model we have applied it to the $\pi$~Mensae system. Constraints on the inclination of $\pi$~Mensae~b from Hipparcos-Gaia astrometry have been derived in several recent works (\citealt{Xuan20}; \citealt{Damasso20}; \citealt{DeRosa20.piMen}), so for our purposes the system serves as a useful point of comparison between different models. As our intention here is to produce indicative rather than authoritative results we have only used a subset of available RV data, namely the AAT dataset used in \citet{Huang18} and \citet{DeRosa20.piMen}, and the re-reduced HARPS RVs of \citet{Trifonov20}. This is similar to the data usage of \citet{Xuan20}, but differs from \citet{DeRosa20.piMen} who used a more extensive HARPS dataset, and especially from \citet{Damasso20} who included RVs from the CORALIE and ESPRESSO spectrographs. Although additional RVs would provide improved parameter constraints, we find that the AAT and HARPS data constrain the RV parameters of $\pi$~Mensae~b to sufficient precision for our purposes. The uncertainties on the astrometric parameters are instead largely driven by the proper motion data.

We provide a comparison of results in Table \ref{table:piMen}. For our model we use the stellar mass of $1.094 \pm 0.039 M_\odot$ from \citet{Huang18}, as in \citet{Xuan20} and \citet{DeRosa20.piMen}, which differs only marginally from the $1.07 \pm 0.04 M_\odot$ adopted by \citet{Damasso20}. Although the combination of RV datasets used by each model differs in every case, all four models broadly agree on the RV parameters of $\pi$~Mensae~b, as an orbital period of $\sim2090$ days, an eccentricity of $\sim0.64$, and an argument of periastron of $\sim330 \degree$ are universally recovered. Though the times of periastron are measured for different epochs, the reported constraints agree well when propagated to the same periastron passage. The values of $K$ reported by \citet{Damasso20} and \citet{DeRosa20.piMen} differ slightly ($\sim2.4 \pm 1.0$ \ms{}), which marginally influences their mass measurements; we find a more similar value to that of \citet{DeRosa20.piMen}, likely due to our use of more similar RV data.

The astrometric parameters show good agreement between the different models. All four sets of parameters find an inclination of $\sim50\degree$, significantly differing from the $i\sim90\degree$ of the transiting $\pi$~Mensae~c \citep{Huang18}. Our values are particularly similar in precision to those of \citet{Xuan20}, whereas \citet{Damasso20} and \citet{DeRosa20.piMen} constrain their astrometric parameters more precisely; however, we note that the stated precisions of \citet{Damasso20} are remarkably high, especially in contrast to those of \citet{DeRosa20.piMen}. While the former uses only the HGCA proper motion data, \citet{DeRosa20.piMen} additionally incorporates the raw Hipparcos astrometry, and would thus be expected to produce more precise results over using the Hipparcos proper motions alone. We are not able to explain this phenomenon, but this does not appear to have affected their nominal values for the astrometric parameters as they agree well with the other results.

Our model is the first to explicitly constrain the barycentric proper motion of $\pi$~Mensae. The derived values listed in Table \ref{table:piMen} differ from the measured proper motions from Gaia DR2 (i.e. $+311.19 \pm 0.13$, $+1048.85 \pm 0.14$ \masyr{}) by $\sim1$ \masyr{}, and furthermore are much better-constrained due in large part to the highly precise Hip-Gaia proper motion measurements.

To conclude, our model finds good agreement with the previously published values for the astrometric orbital parameters of $\pi$~Mensae~b, and additionally provides values for the barycentric proper motion of the system which have not previously been enumerated.

\begin{figure}[t!]
\figurenum{1}
\epsscale{1.15}
\plotone{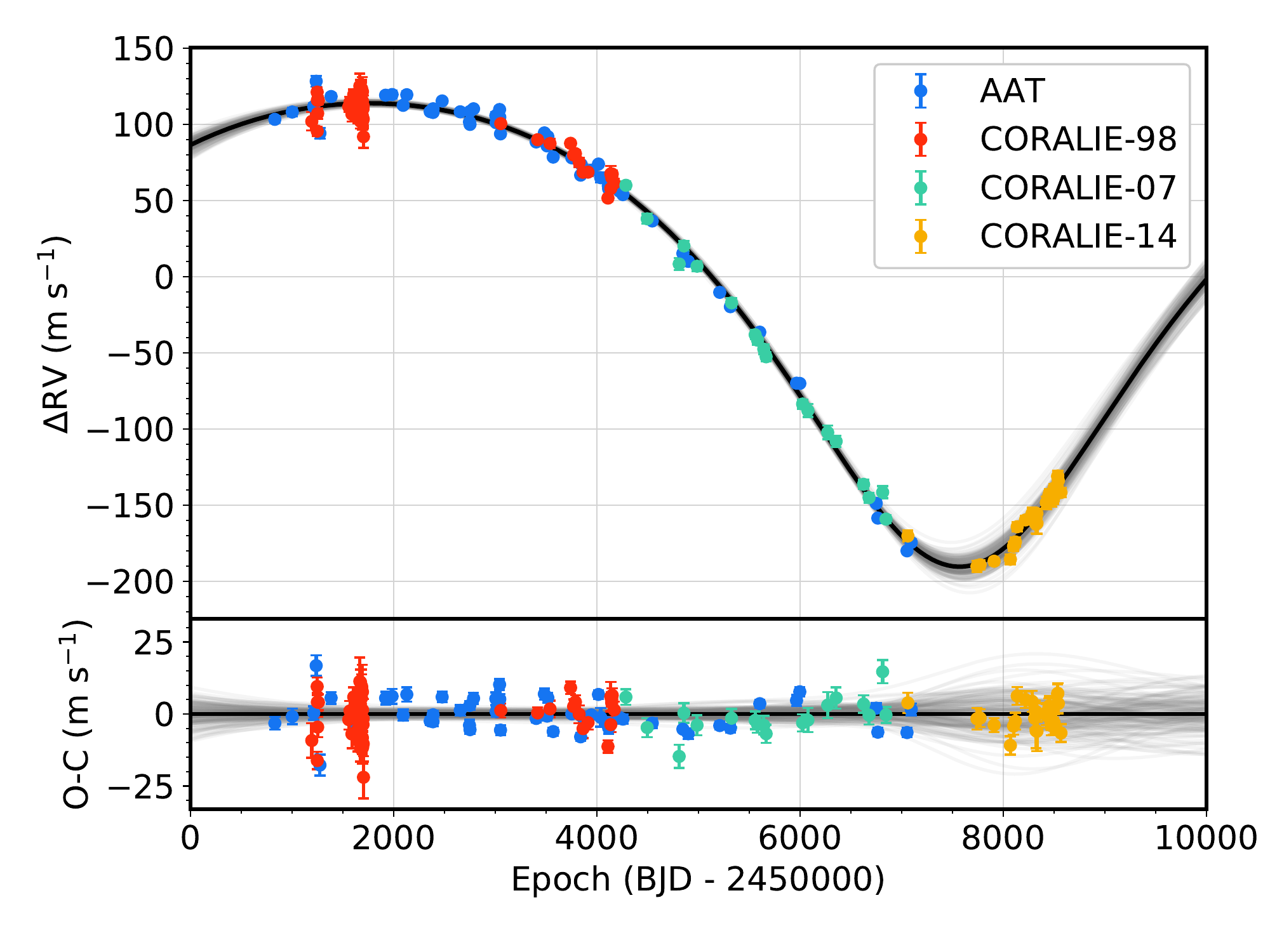}
\caption{RV data and model curves (top) and the residuals (bottom) for HD~92987. The thick black line corresponds to the best-fit parameters whereas the thin grey lines are drawn randomly from the posteriors.
\label{figure:HD92987_RV}}
\end{figure}

\subsection{HD 92987} \label{subsec:results:HD92987}

\begin{deluxetable*}{lccc}[h!]
\centering
\tablecaption{Parameters of HD~92987 B.\label{table:HD92987}}
\tablehead{\colhead{Parameter} & \colhead{\citet{Rickman19}} & \colhead{\citet{Kane19}} & \colhead{This work}}
\startdata
Period $P$ [years] & $28.35^{+1.51}_{-0.74}$ & $29.54^{+2.33}_{-2.19}$ & $31.88^{+0.60}_{-0.53}$ \\
Period $P$ [days] & ($10355^{+550}_{-270}$) & $10790^{+850}_{-800}$ & $11640^{+220}_{-190}$ \\
RV semi-amplitude $K$ [\ms{}] & $152.7^{+2.3}_{-2.7}$ & $162.0^{+14.0}_{-8.8}$ & $152.2 \pm 2.3$ \\
Eccentricity $e$ & $0.21^{+0.02}_{-0.01}$ & $0.25 \pm 0.03$ & $0.252 \pm 0.010$ \\
Argument of periastron (primary) $\omega_1$ [degrees] & $195.1^{+6.7}_{-8.4}$ & $198.4^{+6.9}_{-8.0}$ & $180.2^{+3.1}_{-3.4}$ \\
Time of periastron $T_p$ [JD] & $2457889^{+130}_{-180}$ & -- & $2457579^{+81}_{-89}$ \\
Secondary minimum mass $m_2\sin i$ [$M_J$] & $16.88^{+0.69}_{-0.65}$ & $17.9^{+2.4}_{-1.9}$ & \bmaroon{$17.08 \pm 0.37$} \\
Relative semi-major axis $a$ [AU] $^{(a)}$ & $9.62^{+0.36}_{-0.26}$ & $9.75^{+0.61}_{-0.59}$ & \bmaroon{$10.97^{+0.16}_{-0.14}$} \\
\hline
Orbital inclination $i$ [degrees] & -- & -- & $175.82 \pm 0.07$ \\
Longitude of node (secondary) $\Omega_2$ [degrees] & -- & -- & $74.4^{+1.4}_{-1.3}$ \\
$\sin i$ & -- & -- & $0.0729 \pm 0.0012$ \\
Orbital velocity semi-amplitude $\kappa=\frac{K}{\sin i}$ [\ms{}] & -- & -- & $2087 \pm 19$ \\
Secondary mass $m_2$ [$M_J$] & -- & -- & \bmaroon{$268.4 \pm 4.7$} \\
Secondary mass $m_2$ [${M_\odot}$] & -- & -- & \bmaroon{$0.2562 \pm 0.0045$} \\
\hline
Barycentric RA proper motion $\mu_{bary,RA}$ [\masyr{}] & -- & -- & $+15.50 \pm 0.17$ \\
Barycentric declination proper motion $\mu_{bary,Dec}$ [\masyr{}] & -- & -- & $+10.85 \pm 0.04$ \\
AAT RV offset $\gamma_{\text{AAT}}$ [\ms{}] $^{(b)}$ & -- & -- & $+9.0^{+1.9}_{-1.8}$ \\
Barycentric CORALIE-98 RV $\gamma_{\text{C98}}$ [\ms{}] & -- & -- & $+4770.9^{+2.0}_{-1.9}$ \\
Barycentric CORALIE-07 RV $\gamma_{\text{C07}}$ [\ms{}] & -- & -- & $+4763.2 \pm 2.3$ \\
Barycentric CORALIE-14 RV $\gamma_{\text{C14}}$ [\ms{}] & -- & -- & $+4777.5 \pm 4.2$ \\
\hline
AAT RV jitter $\sigma_{\text{jit,AAT}}$ [\ms{}] & -- & -- & $5.1^{+0.7}_{-0.6}$ \\
CORALIE-98 RV jitter $\sigma_{\text{jit,C98}}$ [\ms{}] & -- & -- & $5.7^{+0.9}_{-0.8}$ \\
CORALIE-07 RV jitter $\sigma_{\text{jit,C07}}$ [\ms{}] & -- & -- & $4.8^{+1.7}_{-1.4}$ \\
CORALIE-14 RV jitter $\sigma_{\text{jit,C14}}$ [\ms{}] & -- & -- & $3.2 \pm 1.0$ \\
\enddata
\tablecomments{(a) Published values for the semi-major axis are based on the minimum mass of HD~92987~B, whereas our value is based on the true mass. (b) As with the absolute radial velocity offsets, this value is \textit{subtracted} from the data in the likelihood calculation.}
\end{deluxetable*}

\begin{figure*}[h!]
\figurenum{2}
\epsscale{1.05}
\plottwo{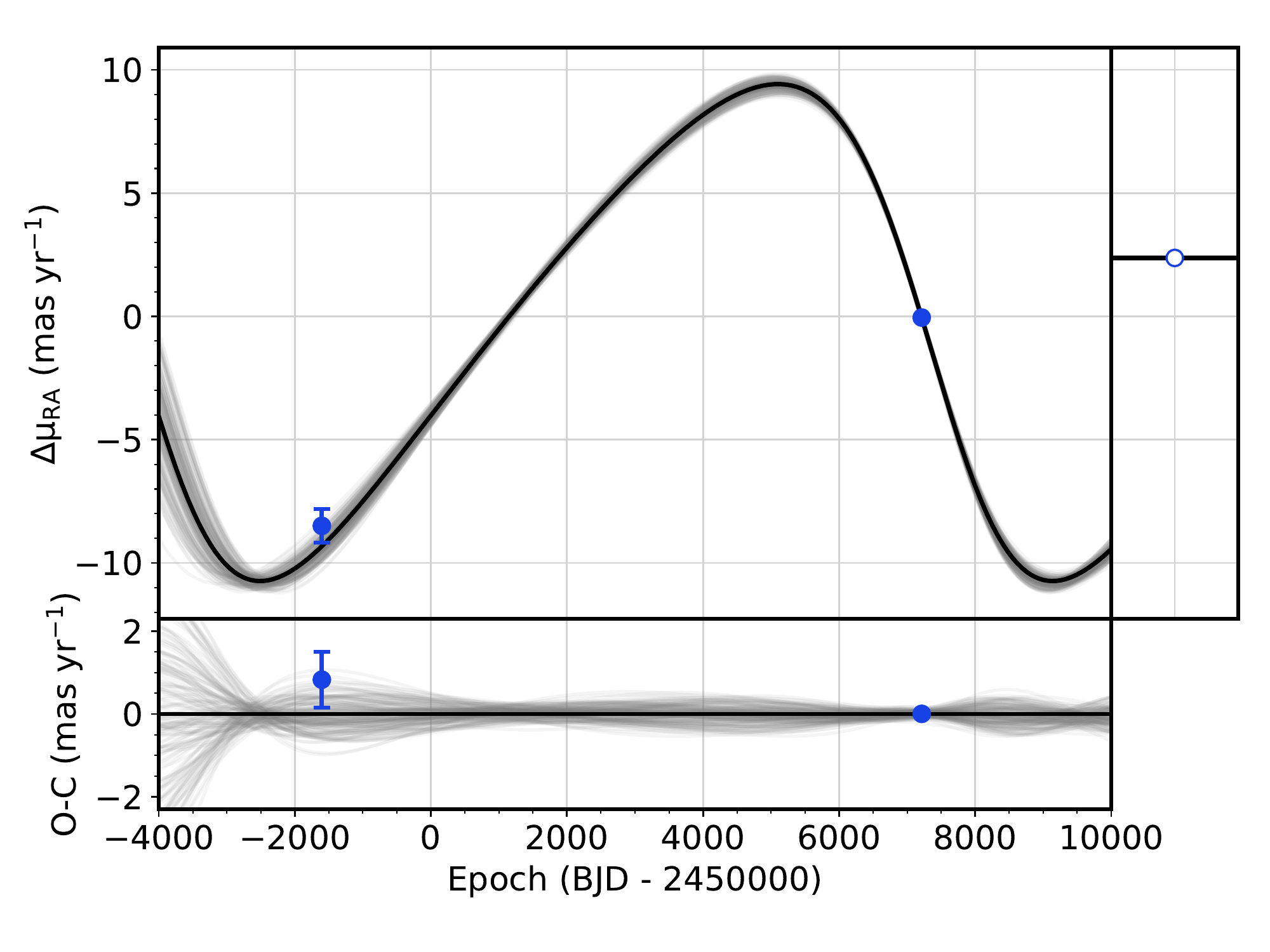}{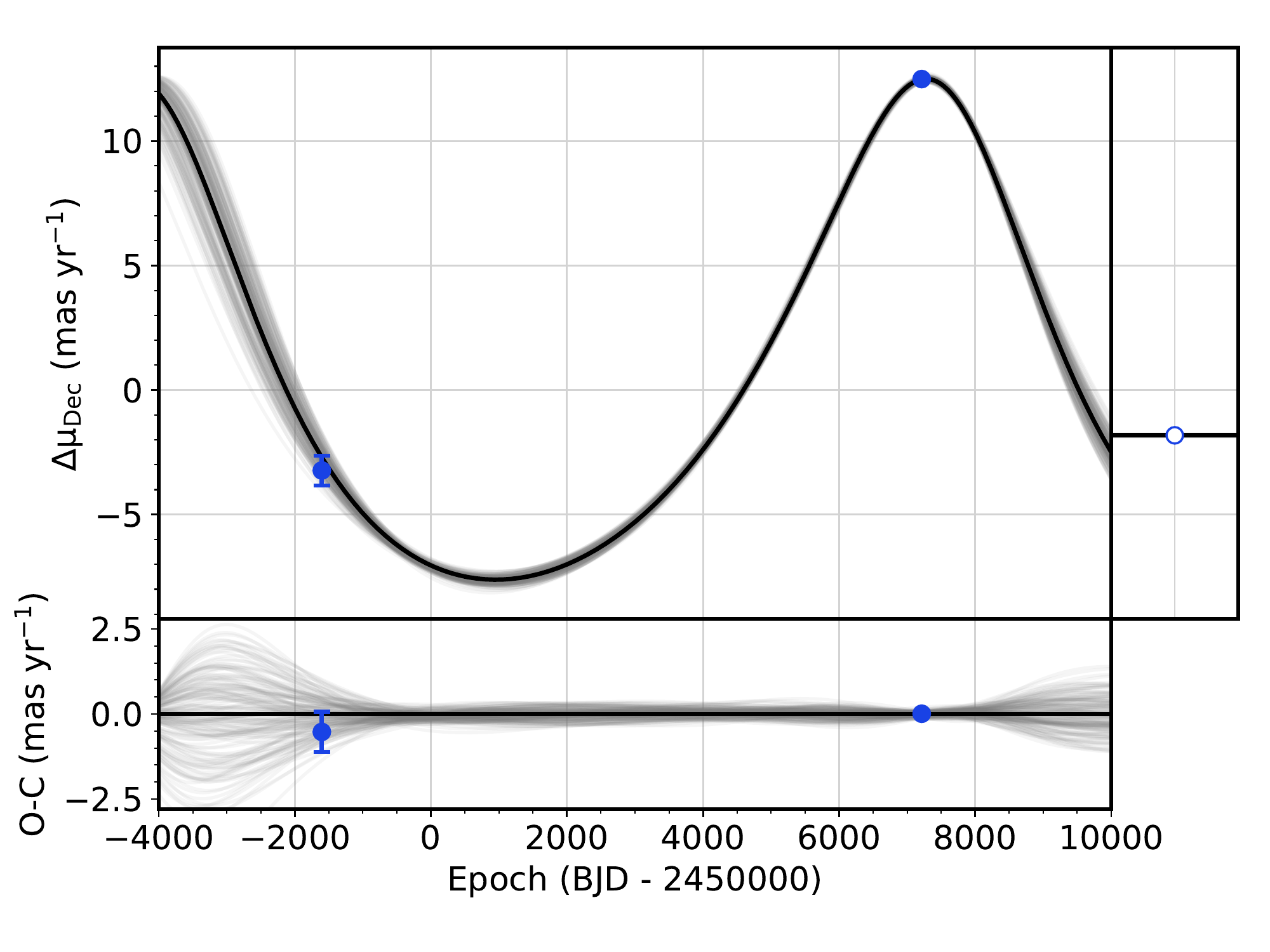}
\caption{Proper motions in right ascension (left) and declination (right) for HD~92987 normalised to the barycentric values. The filled points in the main and residuals panels are the Hipparcos and Gaia measurements, while the unfilled points in the side panels are the Hip-Gaia proper motions. The thick black lines correspond to the best-fit parameters, whereas the thin grey lines are drawn randomly from the posteriors; in the side panel the model line corresponds to the average proper motion between the Hipparcos and Gaia epochs.
\label{figure:HD92987_pm}}
\end{figure*}

\begin{figure}[ht!]
\figurenum{3}
\epsscale{1.15}
\plotone{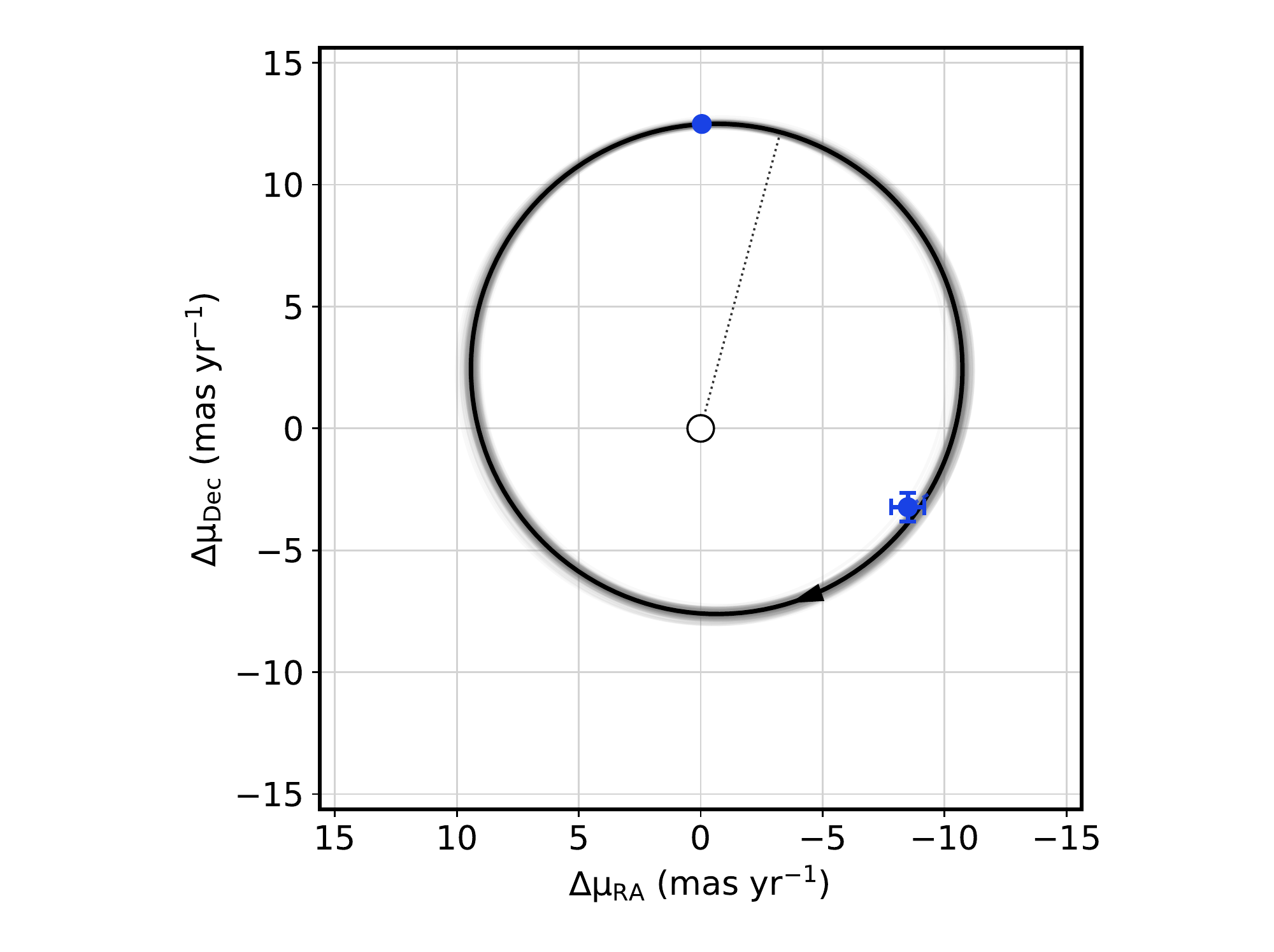}
\caption{Proper motions of HD~92987 illustrated in two dimensions. The thick black line corresponds to the best-fit parameters while the thin grey lines are randomly drawn from the posteriors. The central circle marks the barycentric proper motion while the dotted line connects it to the periastron proper motion, and the arrow demonstrates the direction of motion. Note the reversed X-axis for consistency with the direction of positive right ascension on the sky.
\label{figure:HD92987_pm2D}}
\end{figure}

Having demonstrated the fidelity of our model, we now turn to our two target systems.

In the HGCA (\citealt{Brandt18, Brandt18erratum}) HD~92987 can be easily identified as a star with significant proper motion variability, as previously found by \citet{Frankowski07} and discussed in Section \ref{subsec:targets:HD92987}. The measured difference between the Gaia and Hip-Gaia proper motions is $\sim$14.4 \masyr{}, equivalent to a tangential velocity difference of $\sim$3000 \ms{} for the star's parallax. This is far larger than the amplitude of radial velocity variations, and strongly suggests that the companion is actually a star observed at a near-polar orbital inclination.

Our model confirms this. As shown in Table \ref{table:HD92987}, we find an orbital inclination of $175.82 \pm 0.07 \degree$, only $4.2\degree$ from pole-on, corresponding to $\sin i=0.0729 \pm 0.0012$ and a total orbital velocity semi-amplitude $\kappa=2087 \pm 19$ \ms{}. Combined with the \bmaroon{$17.08 \pm 0.37$} $M_J$ minimum mass the true mass of the companion can be determined as \bmaroon{$268.4 \pm 4.7$} $M_J$ (\bmaroon{$0.2562 \pm 0.0045$} $M_\odot$). This is well above the lower mass limit for hydrogen fusion, so HD~92987~B is confidently identified as a star.

The results for the radial velocity component of our model is shown in Figure \ref{figure:HD92987_RV}, and the results for the astrometric component are shown in Figures \ref{figure:HD92987_pm} and \ref{figure:HD92987_pm2D}. Our values for the RV-constrained parameters are broadly similar to those of \citet{Rickman19} and \citet{Kane19}, which is an unsurprising observation as the radial velocity data used here is a combination of theirs. However, we find that the high signal-to-noise of the proper motion signal results in improved parameter constraints over the RV-only models, with all comparable parameters constrained more precisely. We also find that inclusion of the astrometry leads to a preference for a longer orbital period and smaller argument of periastron at the $\sim$2$\sigma$ level. The degree of improvement for the RV parameters is a special case owing to the strong detection of astrometric variation, as the effect is weaker for HD~221420 (see Section \ref{subsec:results:HD221420}) and negligible for $\pi$~Mensae.

Due to the large proper motion variability induced by HD~92987~B, we find that the barycentric proper motion of the system ($+15.50 \pm 0.17$, $+10.85 \pm 0.04$ \masyr{}) differs by $\sim$9 and $\sim$12 \masyr{} from the Hipparcos and Gaia DR2 proper motions respectively, many times larger than the measurement uncertainties. The uncertainties of the two components of the barycentric proper motion are also markedly asymmetrical, with the uncertainty in right ascension being over 4 times larger than in declination. This is largely a result of the phasing of the Gaia proper motion measurement, which dominates in precision over the Hipparcos measurement but occurs almost precisely at the maximum of $\Delta\mu_{Dec}$ (see Figure \ref{figure:HD92987_pm2D}), therefore providing a much stronger constraint on the amplitude of proper motion variability in declination than in right ascension.

Our model incorporates variable jitter terms for each radial velocity dataset, allowing us to assess the relative non-photonic noise for each instrument. We find that the AAT dataset has an excess jitter of $5.1^{+0.7}_{-0.6}$ \ms{} whereas the CORALIE datasets have jitters of $5.7^{+0.9}_{-0.8}$, $4.8^{+1.7}_{-1.4}$, and $3.2 \pm 1.0$ \ms{} respectively. For CORALIE this suggests decreasing levels of noise with each successive upgrade to the spectrograph, although the uncertainty on the CORALIE-07 jitter parameter is too large to infer this with confidence.

\begin{figure*}[p!]
\figurenum{4}
\epsscale{1.2}
\plotone{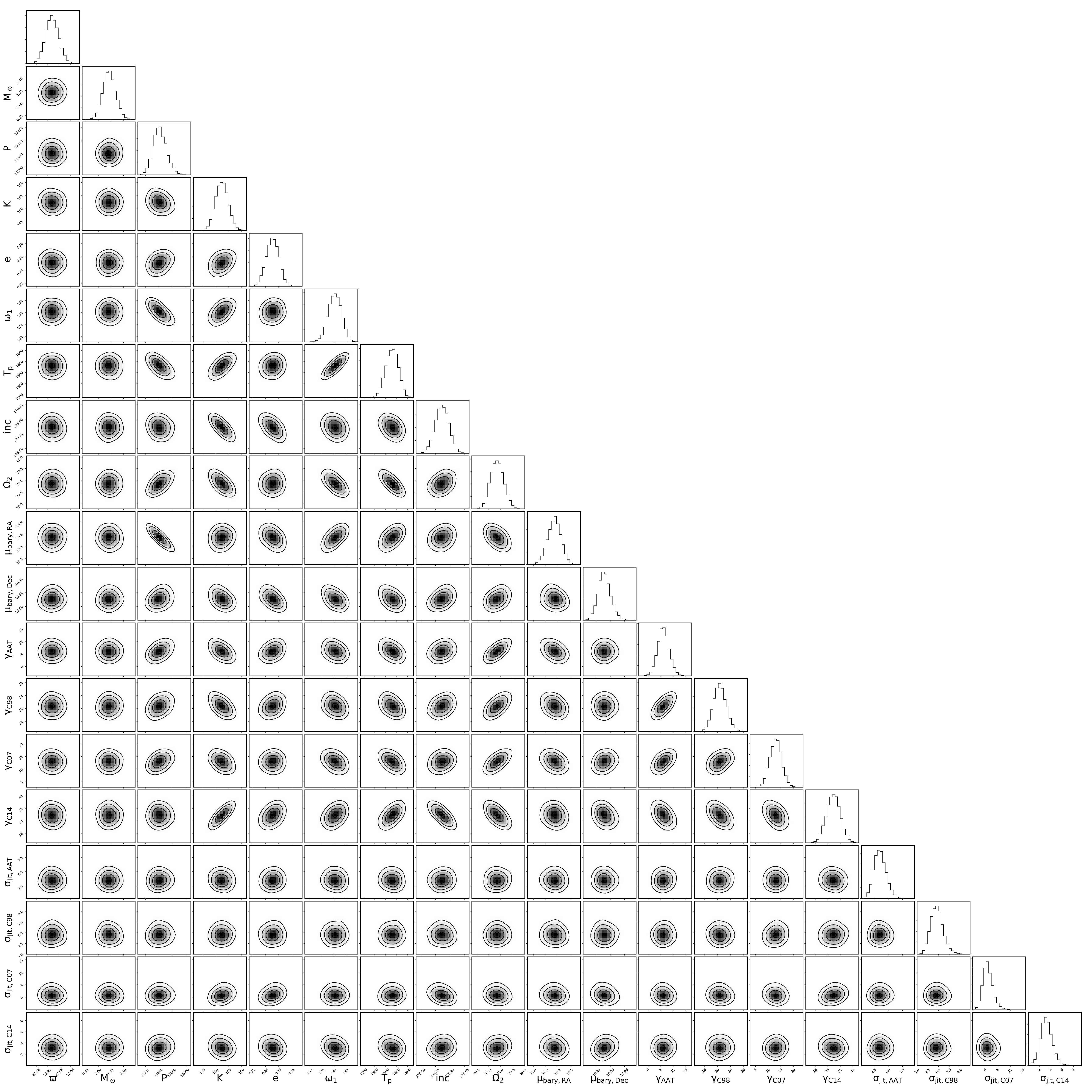}
\caption{Joint posterior distributions and histograms of the posteriors for each variable parameter used in the joint model of the HD~92987 system. Parameters are as in Section \ref{subsec:modelfit}. An offset of JD 2450000 has been subtracted from $T_p$ and an offset of $+4750$ \ms{} has been subtracted from the $\gamma_{\text{CORALIE}}$ parameters for clarity.
\label{figure:HD92987_corner}}
\end{figure*}

The joint posterior distributions and histograms for our model of HD~92987 are shown in Figure \ref{figure:HD92987_corner}. All parameters appear well behaved and approximately Gaussian in distribution. A handful of parameter relationships explain most of the observed correlations and these can be easily explained; $\omega_1$ and $T_p$ are naturally strongly related as they both relate to periastron; those two parameters are then correlated with $P$ as this partially controls the timing of periastron; $\mu_{bary,Dec}$ displays an especially strong correlation with $P$ due to the phasing of the Gaia measurement as discussed previously; finally, \textbf{$\gamma_{\text{C14}}$} is strongly correlated with $K$ because the CORALIE-14 dataset can only narrowly be compared with earlier RV datasets, so the best-fit offset for that data is significantly dependent on the RV semi-amplitude (this is visible in Figure \ref{figure:HD92987_RV}).

In summary, we find similar RV parameters for HD~92987~B as in previous works, but the astrometric data leads us to discover that the companion is a \bmaroon{$0.2562 \pm 0.0045$} $M_\odot$ star observed at a near-polar orbital inclination rather than a substellar object.

\subsection{HD 221420} \label{subsec:results:HD221420}

\begin{figure}[t!]
\figurenum{5}
\epsscale{1.15}
\plotone{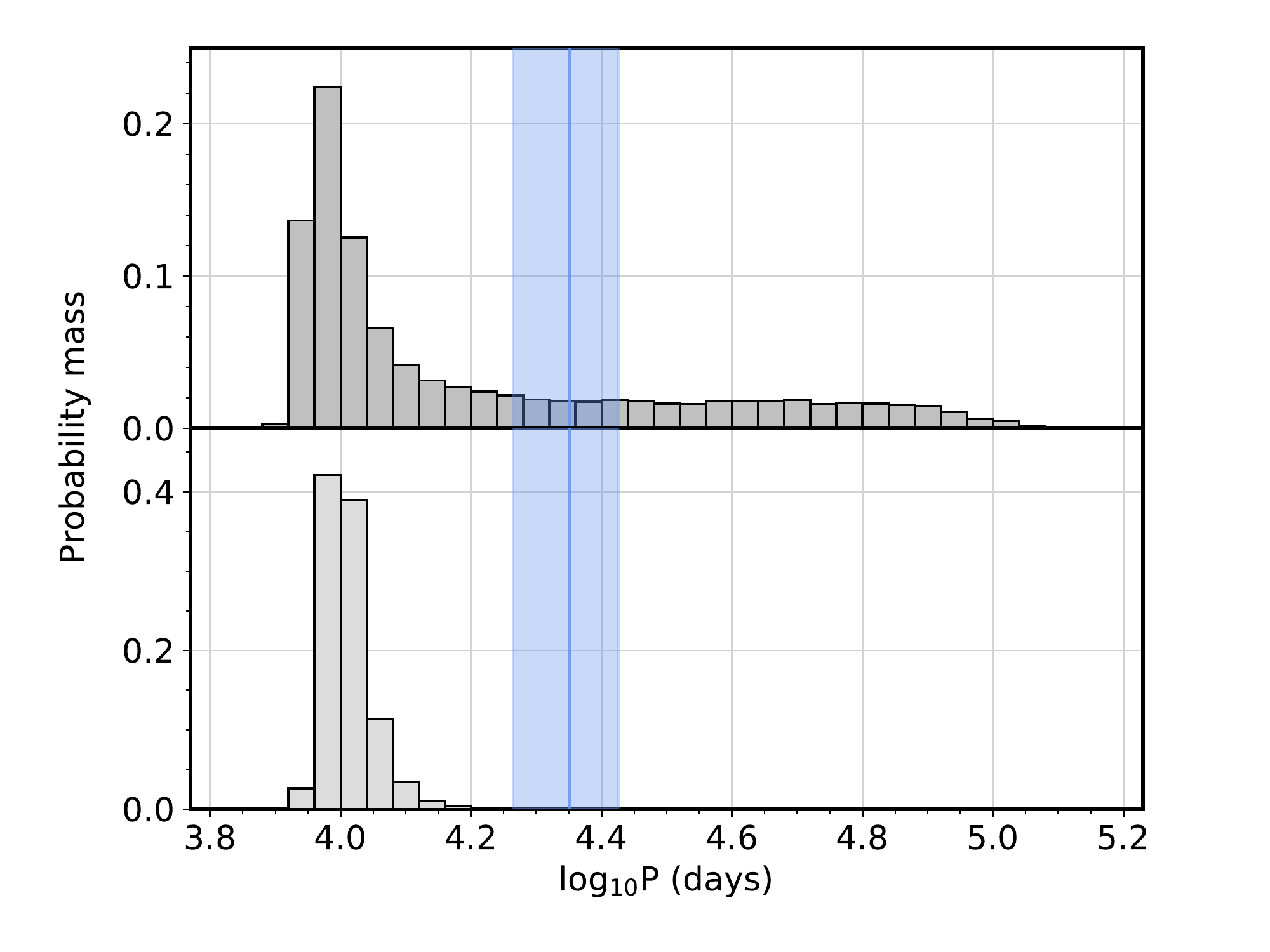}
\caption{Distribution of orbital periods for HD~221420~b using radial velocities only (top) and using RVs + proper motions (bottom). The orbital period of $22482^{+4200}_{-4100}$ days reported by \citet{Kane19} is shown in blue. Our RV-only model shows a preference for orbital periods of $\sim$9200 days, which may differ from the previous period due to the inclusion of the HARPS RVs, but there is a long tail towards longer orbital periods such that the period from \citet{Kane19} cannot be rejected. Inclusion of the proper motion data is sufficient to reject these longer orbital periods, resulting in an unambiguous shorter period of $10090^{+890}_{-560}$ days.
\label{figure:HD221420_Pdist}}
\end{figure}

In the HGCA, HD~221420 has a moderately large astrometric acceleration; the measured Gaia-HG proper motion anomaly is $\sim$1.5 \masyr{}, equivalent to a tangential velocity difference of $\sim$220 \ms{}. This is larger than the amplitude of the radial velocity variations, but not to as great a degree as for HD~92987.

In our model we find a distinctly shorter orbital period than \citet{Kane19}, which merits explanation. In Figure \ref{figure:HD221420_Pdist} we plot the posterior distribution of orbital periods for our joint model as well as for a RV-only model that excludes proper motion data. The RV-only model displays a preference for orbital periods around $\sim$9200 days, possibly caused by the novel inclusion of HARPS data; however, the distribution has a tail towards longer orbital periods that do not provide significantly worse fits, and we would not be able to reject the orbital period of \citet{Kane19} based on RVs alone. Inclusion of the longer-spanning astrometric data resolves this uncertainty with a shorter orbital period being supported by the similarity between the Hipparcos and Gaia proper motions, which implies that the true orbital period is similar to the $\sim$24 year ($\sim$8800 day) interval between the two measurements.

\begin{figure}[t!]
\figurenum{6}
\epsscale{1.15}
\plotone{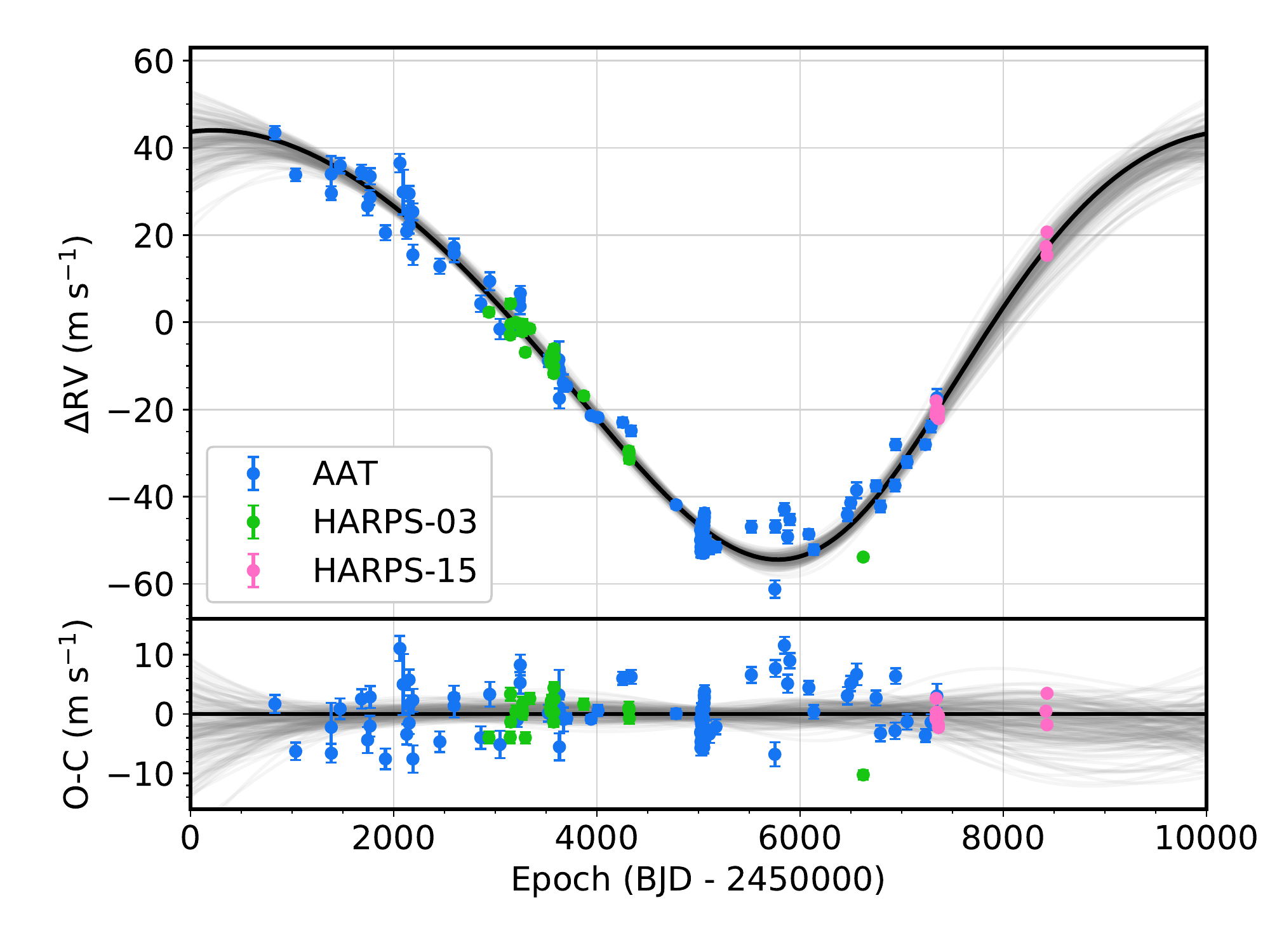}
\caption{RV data and model curves (top) and the residuals (bottom) for HD~221420. The thick black line corresponds to the best-fit parameters whereas the thin grey lines are randomly drawn from the posteriors.
\label{figure:HD221420_RV}}
\end{figure}

\begin{figure*}[h!]
\figurenum{7}
\epsscale{1.05}
\plottwo{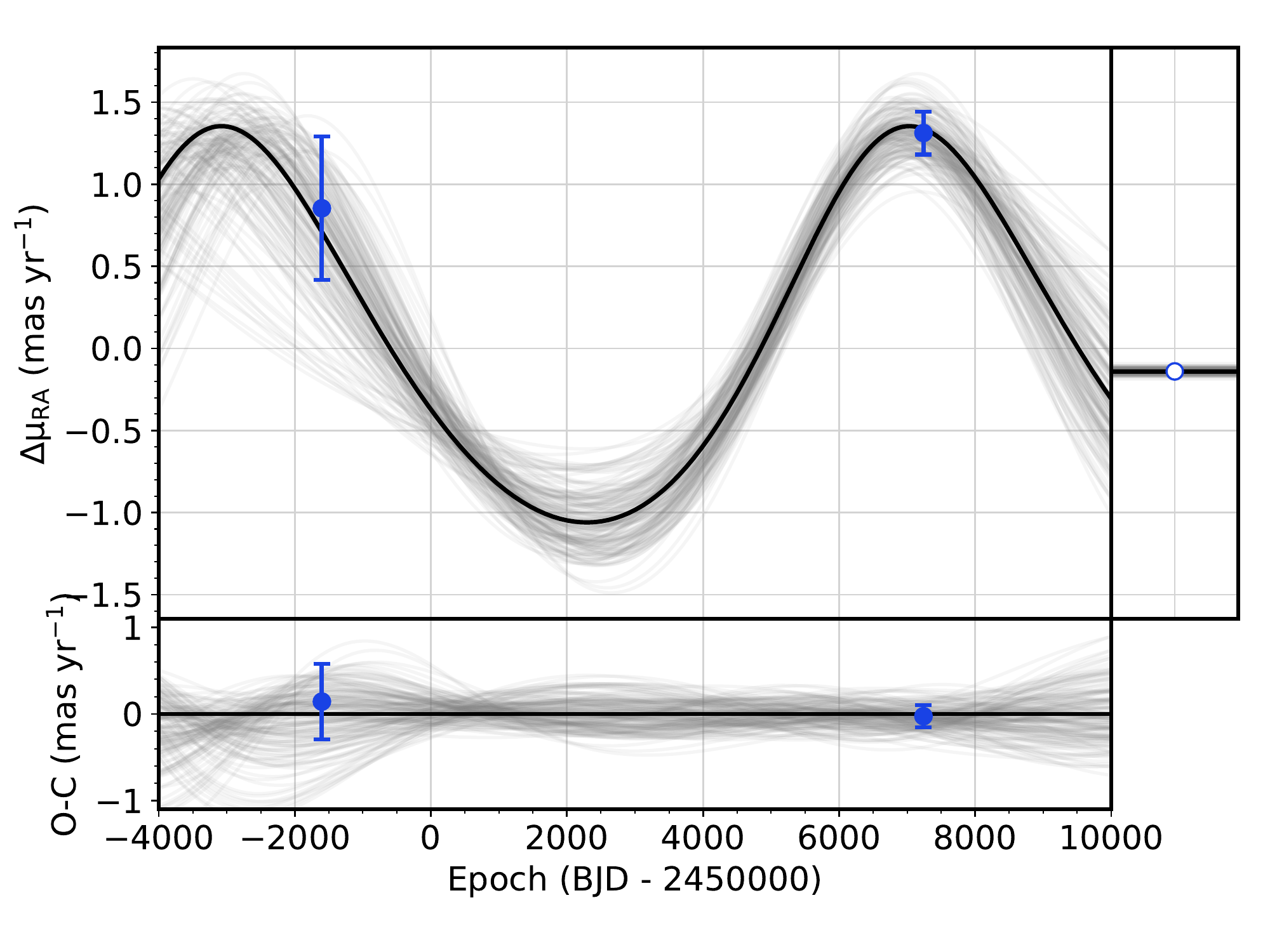}{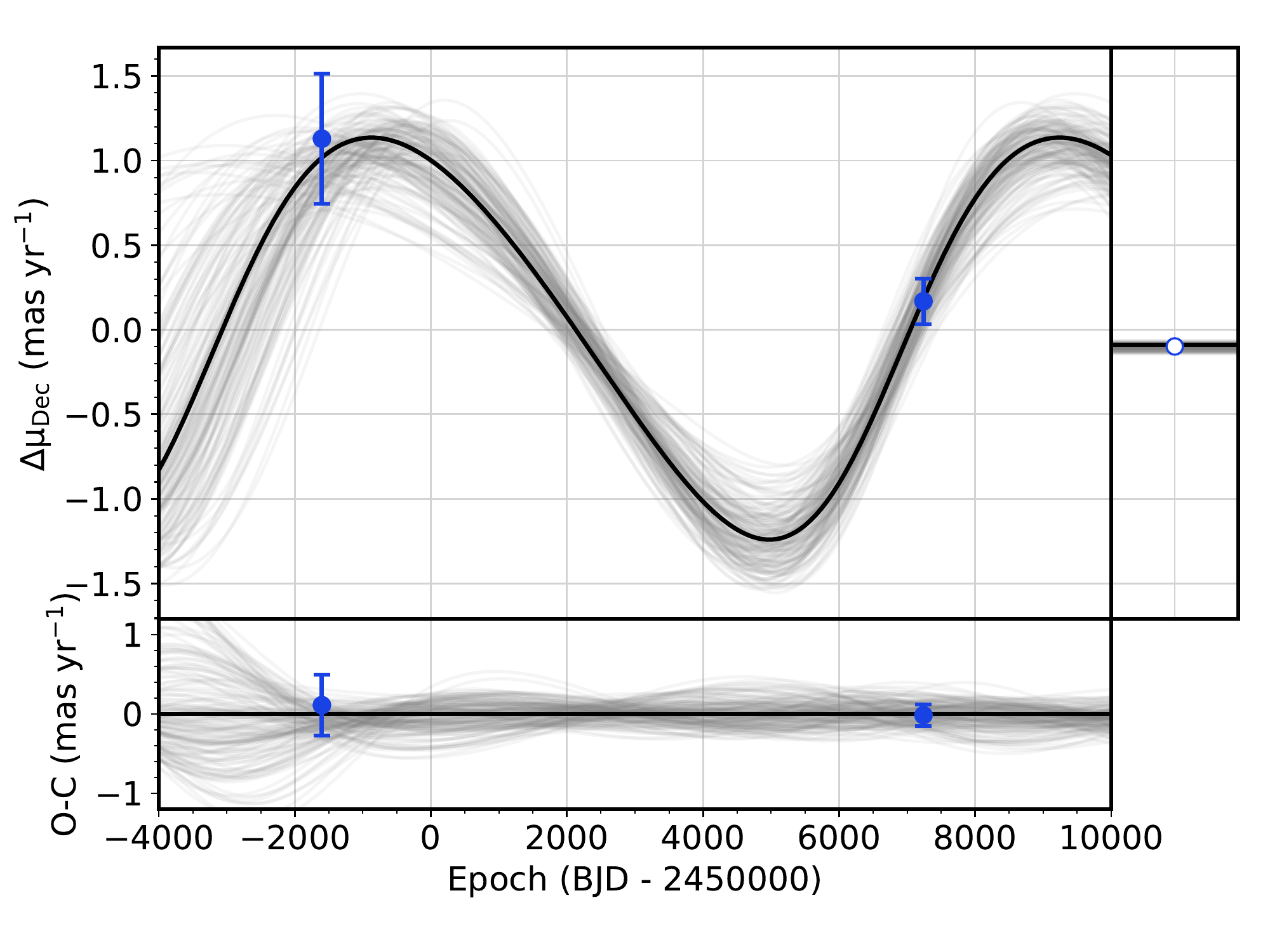}
\caption{Proper motions in right ascension (left) and declination (right) for HD~92987 normalised to the barycentric values. As in Figure \ref{figure:HD92987_pm}, the filled points are the Hipparcos and Gaia measurements while the unfilled points are the Hip-Gaia proper motions, and the black lines correspond to the best-fit parameters while the thin grey lines are drawn randomly from the posteriors. As discussed in the text long orbital periods for HD~221420~b tend to underestimate $\Delta\mu_{RA}$ at the Hipparcos epoch, and it is this that drives the preference for relatively shorter periods in our model.
\label{figure:HD221420_pm}}
\end{figure*}

\begin{figure}[h!]
\figurenum{8}
\epsscale{1.15}
\plotone{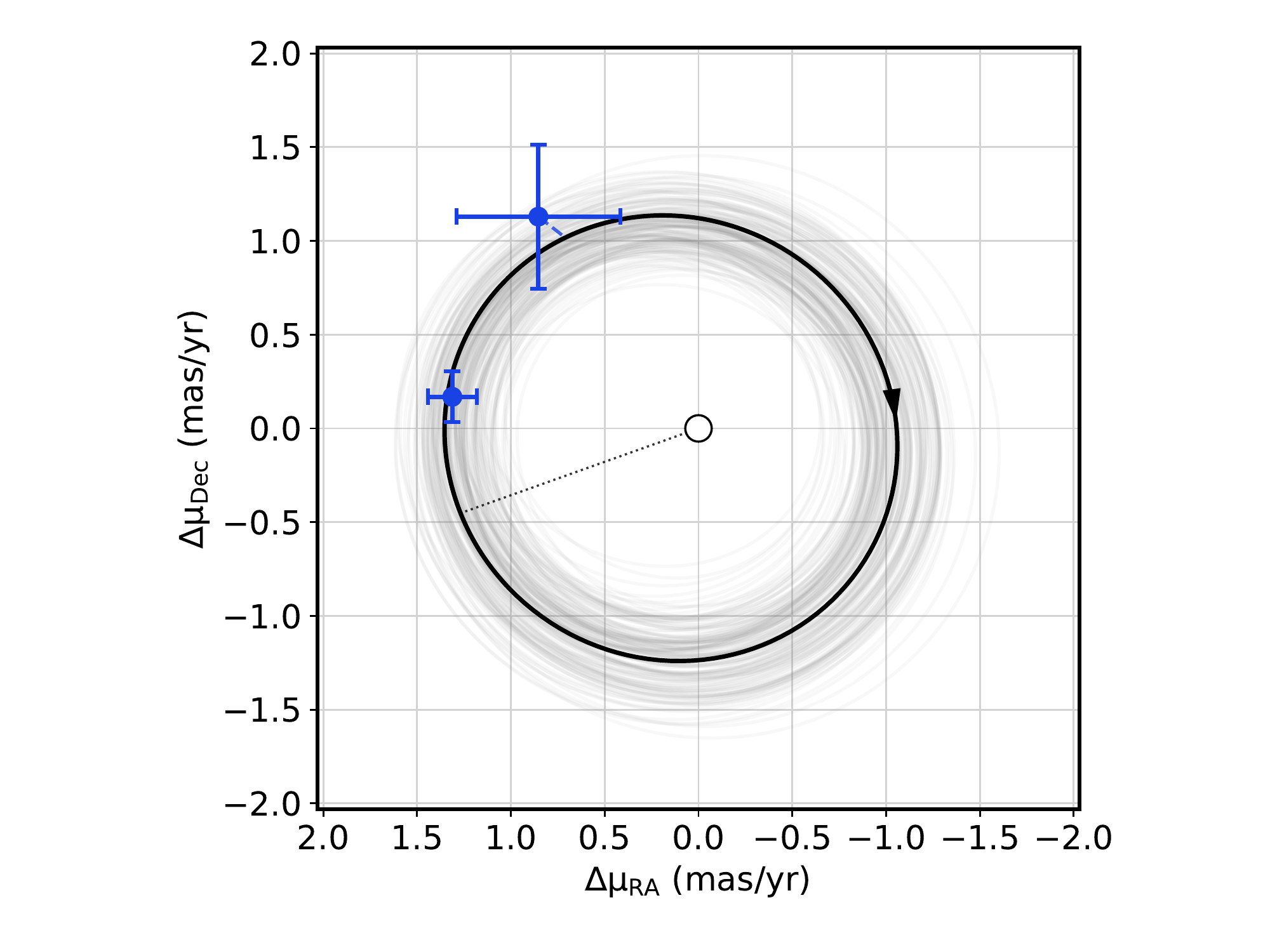}
\caption{Proper motions of HD~221420 illustrated in two dimensions. As in Figure \ref{figure:HD92987_pm2D}, the thick black line corresponds to the best-fit parameters whereas the thin grey lines are randomly drawn from the posteriors, the central circle marks the barycentric proper motion while the dotted line connects it to the periastron proper motion, and the arrow shows the direction of motion. The similar orbital phasing of the Hipparcos and Gaia measurements provides relatively strong constraints on the orbital period of HD~221420~b.
\label{figure:HD221420_pm2D}}
\end{figure}

The results of our model for HD~221420 are listed in Table \ref{table:HD221420}, with the radial velocity component shown in Figure \ref{figure:HD221420_RV} and the astrometric component in Figures \ref{figure:HD221420_pm} and \ref{figure:HD221420_pm2D}. The aforementioned similarity between $\Delta\mu_H$ and $\Delta\mu_G$ is visible in the latter figure, and our value for the orbital period of $27.62^{+2.45}_{-1.54}$ years ($10090^{+890}_{-560}$ days) is indeed comparable to the interval between Hipparcos and Gaia observations. Orbital periods closer to those of \citet{Kane19} tend to underestimate $\Delta\mu_{RA}$ at the Hipparcos epoch, which is visible in Figure \ref{figure:HD221420_pm}. Using the astrometric data we find an orbital inclination of $164.0^{+1.9}_{-2.6}$ $\degree$, corresponding to $\sin i=0.276^{+0.044}_{-0.032}$ and $\kappa=176 \pm 18$ \ms{}. From a minimum mass of $6.3^{+0.5}_{-0.3}$ $M_J$ we thus find a true companion mass of $22.9 \pm 2.2$ $M_J$.

Our values for the barycentric proper motion of the system, $+15.00 \pm 0.06$ and $+0.65^{+0.08}_{-0.05}$ \masyr{}, differ by $\sim$1.4 and $\sim$1.3 \masyr{} from the Hipparcos and Gaia DR2 values respectively. Unlike for HD~92987 the uncertainties for the two components are relatively even, which we attribute to differences in observational phasing and the relatively smaller disparity between Hipparcos and Gaia measurement precision. In Appendix \ref{appendix:A} we use these values for the barycentric proper motion of HD~221420 to investigate the possibility of its association with a nearby star.

For the jitter parameters, we find values of $4.0 \pm 0.4$ \ms{} for the AAT dataset and $3.0^{+0.6}_{-0.5}$ and $2.4^{+1.1}_{-0.6}$ \ms{} for the two HARPS datasets respectively. The AAT jitter is similar to that found for HD~92987, while the pre- and post-upgrade HARPS jitter parameters are compatible within the uncertainties.

\begin{deluxetable*}{lcc}[h!]
\centering
\tablecaption{Parameters of HD 221420 b.\label{table:HD221420}}
\tablehead{\colhead{Parameter} & \colhead{\citet{Kane19}} & \colhead{This work}}
\startdata
Period $P$ [years] & $61.55^{+11.50}_{-11.23}$ & $27.62^{+2.45}_{-1.54}$ \\
Period $P$ [days] & $22482^{+4200}_{-4100}$ & $10090^{+890}_{-560}$ \\
RV semi-amplitude $K$ [\ms{}] & $54.7^{+4.2}_{-3.6}$ & $48.5^{+2.5}_{-1.9}$ \\
Eccentricity $e$ & $0.42^{+0.05}_{-0.07}$ & $0.14^{+0.04}_{-0.03}$ \\
Argument of periastron (primary) $\omega_1$ [degrees] & $164.4^{+6.9}_{-6.3}$ & $211^{+10}_{-11}$ \\
Time of periastron $T_p$ [JD] & -- & $2456440^{+230}_{-260}$ \\
Secondary minimum mass $m_2\sin i$ [$M_J$] & $9.7^{+1.1}_{-1.0}$ & $6.3^{+0.5}_{-0.3}$ \\
Relative semi-major axis $a$ [AU] & $18.5^{+2.3}_{-2.3}$ & $10.15^{+0.59}_{-0.38}$ \\
\hline
Orbital inclination $i$ [degrees] & -- & $164.0^{+1.9}_{-2.6}$ \\
Longitude of node (secondary) $\Omega_2$ [degrees] & -- & $236.9^{+5.7}_{-5.6}$ \\
$\sin i$ & -- & $0.276^{+0.044}_{-0.032}$ \\
Orbital velocity semi-amplitude $\kappa=\frac{K}{\sin i}$ [\ms{}] & -- & $176 \pm 18$ \\
Secondary mass $m_2$ [$M_J$] & -- & $22.9 \pm 2.2$ \\
\hline
Barycentric RA proper motion $\mu_{bary,RA}$ [\masyr{}] & -- & $+15.00 \pm 0.06$ \\
Barycentric declination proper motion $\mu_{bary,Dec}$ [\masyr{}] & -- & $+0.65^{+0.08}_{-0.05}$ \\
AAT RV offset $\gamma_{\text{AAT}}$ [\ms{}] $^{(a)}$ & -- & $+43.3^{+4.2}_{-2.8}$ \\
HARPS-03 RV offset $\gamma_{\text{H03}}$ [\ms{}] $^{(a,}$ $^{b)}$ & -- & $+10.3^{+4.2}_{-3.0}$ \\
Barycentric HARPS-15 RV $\gamma_{\text{H15}}$ [\ms{}] & -- & $+26590.1^{+4.5}_{-3.3}$ \\
\hline
AAT RV jitter $\sigma_{\text{jit,AAT}}$ [\ms{}] & -- &  $4.0 \pm 0.4$ \\
HARPS-03 RV jitter $\sigma_{\text{jit,H03}}$ [\ms{}] & -- &  $3.0^{+0.6}_{-0.5}$ \\
HARPS-15 RV jitter $\sigma_{\text{jit,H15}}$ [\ms{}] & -- &  $2.4^{+1.1}_{-0.6}$ \\
\enddata
\tablecomments{(a) As in Table \ref{table:HD92987}, this value is subtracted from the data in the likelihood calculation. (b) Although the HARPS-03 radial velocities would normally be absolute the re-reduction of \citet{Trifonov20} normalises the data to a constant velocity, so the offset given here is relative.}
\end{deluxetable*}

\begin{figure*}[p!]
\figurenum{9}
\epsscale{1.2}
\plotone{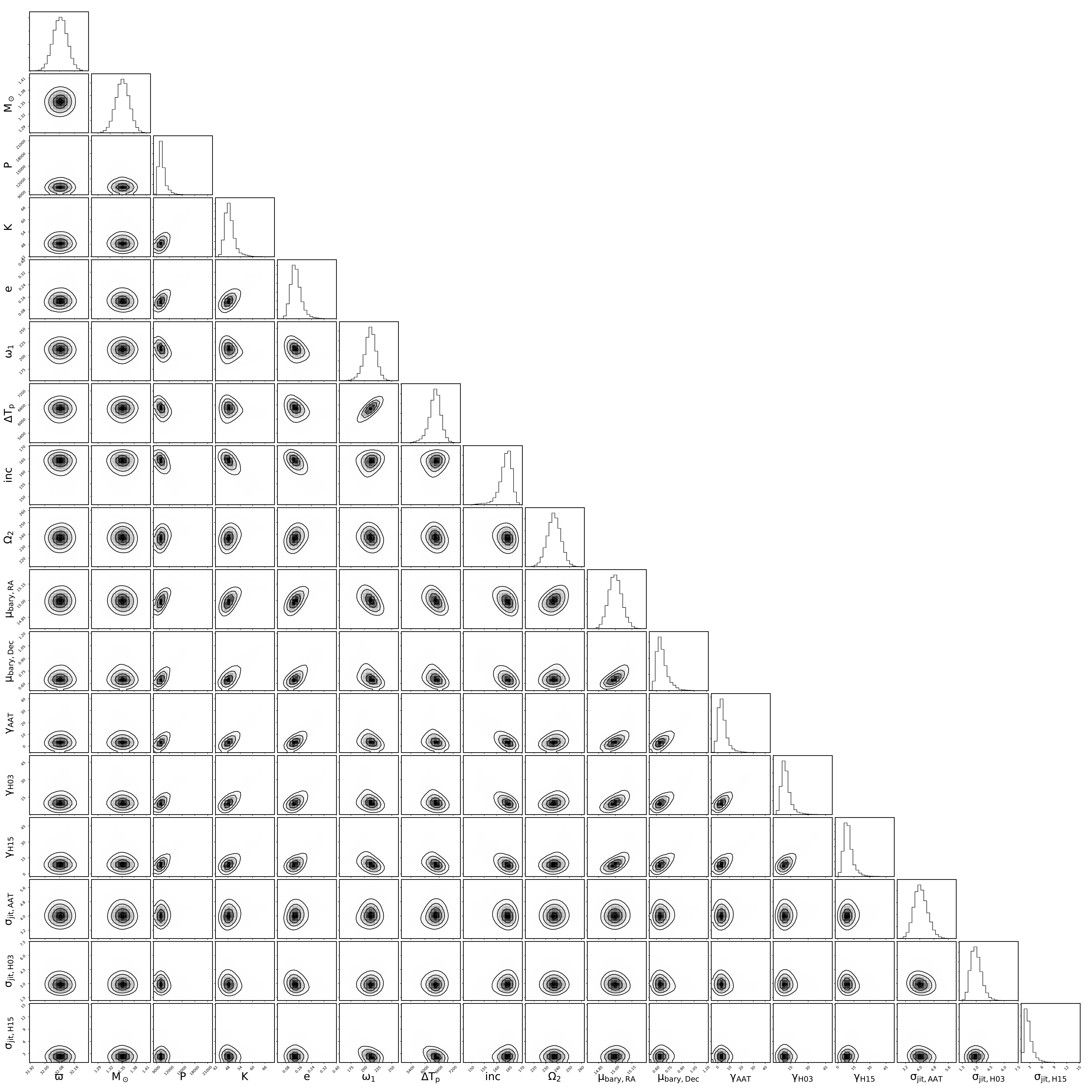}
\caption{Joint posterior distributions and histograms for HD~221420. Parameters are as in Section \ref{subsec:modelfit}. An offset of JD 2450000 has been subtracted from $T_p$ and an offset of $+26580$ \ms{} has been subtracted from $\gamma_{\text{H15}}$ for clarity.
\label{figure:HD221420_corner}}
\end{figure*}

The joint posterior distributions and histograms for HD~221420 are shown in Figure \ref{figure:HD221420_corner}. Compared to HD~92987 (Figure \ref{figure:HD92987_corner}) the parameters are not as well behaved; many of the parameter distributions are decidedly \bmaroon{skewed and} asymmetrical, which appears to be driven by the unequal distribution of $P$. All parameters except the jitters display some degree of correlation with $P$ and $K$, \bmaroon{including} all of the $\gamma$ parameters, which suggests that the available radial velocity data is insufficient to constrain the RV offsets unambiguously. Further RV observations \bmaroon{and astrometry} will be necessary to reduce the parameter correlations observed here.

To summarise, our joint model for HD~221420 finds a shorter orbital period for the companion than previous results, with corresponding reductions in the minimum mass and eccentricity. From the proper motion data we find a true mass for HD~221420~b of $22.9 \pm 2.2$ $M_J$, which is around four times larger than the minimum mass but is still well within the substellar range.

\section{Discussion} \label{sec:discussion}

\subsection[HD 92987 B, hidden star at low inclination]{HD 92987 B, a star masquerading as a substellar object}

In our results we have found that HD~92987~B lies at an orbital inclination of $175.82 \pm 0.07\degree$, only $4.2\degree$ from pole-on. While the minimum companion mass is a nominally substellar \bmaroon{$17.08 \pm 0.37$} $M_J$, the true mass is actually \bmaroon{$0.2562 \pm 0.0045$} $M_\odot$, rendering it unambiguously stellar. HD~92987~B therefore joins the list of apparent substellar companions discovered using RVs that are in fact stellar, such as HD~33636~B ($m\sin i=9.3$ $M_J$, $m=0.14\pm0.01$ $M_\odot$; \citealt{Bean07}), HD~43848~B ($m\sin i=25$ $M_J$, $m=0.10\pm0.01$ $M_\odot$; \citealt{Sozzetti10, Sahlmann11}), HD~211847~B ($m\sin i\approx20$ $M_J$, $m=0.148\pm0.008$ $M_\odot$; \citealt{Moutou17}), HD~202206~B ($m\sin i=17.4$ $M_J$, $m=0.089^{+0.007}_{-0.006}$ $M_\odot$; \citealt{Benedict17.202206}), and various objects from \citet{Sahlmann11}, \citet{Wilson16}, and \citet{Kiefer19}. Although the chance alignment of a binary's orbital pole with our line of sight in such a manner as to produce an apparent low-mass substellar object is relatively improbable, a disproportionately large share of candidate brown dwarf companions are actually stars because genuine brown dwarf companions are infrequent compared to stellar binaries, a phenomenon known as the brown dwarf desert \citep{Grether06}.

The large amplitude of the astrometric signal observed for HD~92987 means that the stellar nature of the companion could still be identified even if the signal-to-noise ratio were several times lower. This suggests that Hipparcos-Gaia astrometry could be used to efficiently determine true masses for similar substellar companion candidates, especially those with periods that are too long to be reliably constrained from Hipparcos or Gaia astrometry alone. In this context the more modest astrometric amplitude of HD~221420 is informative, as it is an example of a companion with similar orbital parameters but an order of magnitude lower true mass. Though decidedly less significant than for HD~92987, the proper motion variability of HD~221420 is still large enough to be used to confidently identify its true mass. As a result, we argue that Hipparcos-Gaia astrometry is a valuable means for determining the true mass of long-period substellar companion candidates, and especially for determining whether these are actually stars observed at low orbital inclination.

\bmaroon{We note that while the Gaia DR2 solution for HD~92987 reports no astrometric excess noise and a low Renormalised Unit Weight Error (RUWE) of 1.17 for the star, the recently released Gaia EDR3 solution \citep{GaiaEDR3} includes a significant excess noise of 0.36 mas and a RUWE of 2.05. A RUWE significantly above 1 indicates a poor fit to the Gaia astrometry \citep{GaiaDR2}, so this suggests that a single-star model is an increasingly poor descriptor of the motion of HD~92987. Considering the 12 month extension in data used for the EDR3 astrometric solution, it appears likely that the increased excess noise and RUWE in Gaia EDR3 is a result of the slow acceleration of HD~92987~A by HD~92987~B. The full Gaia DR3 solution is expected to include non-linear models for source motion, which should provide a direct measurement of astrometric acceleration of HD~92987 that could be used to provide improved constraints on the orbit of its companion.}

\subsection[HD 221420 b, planet or BD?]{HD 221420 b, brown dwarf or ``super-planet"?}

Our model finds an orbital inclination for HD~221420~b of $164.0^{+1.9}_{-2.6}$ $\degree$ and a true mass of $22.9 \pm 2.2$ $M_J$. From this mass the companion is unambiguously substellar, but whether or not it should be referred to as a planet is a matter for debate.

The IAU working definition for an exoplanet considers substellar objects with true masses below 13 $M_J$ (taken to be the upper mass limit for deuterium fusion) to be planets, whereas substellar objects above this limit are brown dwarfs (\citealt{Boss03, Boss07}). Under this definition, HD~221420~b would be considered a brown dwarf.

The IAU working definition is used widely, but not unanimously. Some past authors have argued that the strict dependence on mass is inadequate and have instead advocated for a definition based on formation mechanism, where planets are substellar objects which form via core accretion and brown dwarfs are those which form directly via gravitational collapse (see \citealt{Schlaufman18} for a summary). Categorisation of substellar objects in this way is appealing, but it is not currently applicable in practice as it is largely impossible to determine the formation mechanism of individual objects. However, it may be possible to infer aspects of the distribution of formation mechanisms for the overall population based on the aggregate of companion parameters, a possibility which has been taken up by several recent works.

A correlation between stellar metallicity and the occurrence rate of giant planets, commonly known as the planet-metallicity correlation, has been recognised since the early days of exoplanetology (\citealt{Gonzalez97}; \citealt{Fischer05}). This relationship is commonly interpreted as a result of planet formation by core accretion, which preferentially favours metal-rich protoplanetary disks (\citealt{Hasegawa14}; \citealt{Adibekyan19}). In two similar works, \citet{Santos17} and \citet{Schlaufman18} investigated this relationship further by considering the role of planetary mass and found evidence for differing metallicity dependence for different companion masses, with planets less massive than 4 $M_J$ displaying a preference for high metallicities and more massive planets showing a metallicity distribution closer to that of field stars. \citet{Goda19} performed a comparable study additionally considering stellar mass and planetary eccentricity as variables, as well as making more stringent corrections for sample biases; they found that the companion mass-metallicity dependence is further dependent on stellar mass, with planets orbiting solar-mass stars between ($0.8-1.3$ ${M_\odot}$) showing a consistent high metallicity preference up to $\sim25$ $M_J$, but planets orbiting more massive stars showing a less skewed metallicity distribution, with those above 4 $M_J$ even preferring metal-poor stars. Differences in the eccentricity distribution are also pronounced, with more massive companions having higher average eccentricities regardless of host mass. These results are suggestive of different populations in the massive planet regime, which appear to relate to different formation mechanisms. \citet{Goda19} particularly stress the importance of a ``hybrid scenario" for giant planet formation, with multiple formation mechanisms producing overlapping populations of substellar companions.

Recently, \citet{Bowler20} investigated the eccentricity distribution of directly imaged substellar companions. Although the model orbits of these objects are typically based on short arcs and thus have high uncertainties, the authors were able to extract the eccentricity distribution of the population ensemble using Hierarchical Bayesian modelling. They found that the eccentricity distribution for directly imaged companions differs significantly between low and high-mass objects, with companions below their adopted division of 15 $M_J$ preferring eccentricities below $\sim0.25$ but more massive companions having a much higher average eccentricity ($\sim0.6-0.9$), resembling stellar binaries. This result is similar to that found for eccentricities in \citet{Goda19}, and together they suggest a relationship between eccentricity and formation mechanism.

In light of these past studies, the formation mechanism behind HD~221420~b can be considered. With a mass of $22.9 \pm 2.2$ $M_J$, HD~221420~b lies near to the $\sim$25 $M_J$ upper limit for objects that can plausibly be produced via core accretion according to \citet{Goda19}. The high host star Fe/H of $0.33 \pm 0.02$ makes this more plausible, but it is also entirely conceivable that the companion formed via other mechanisms. The low orbital eccentricity of $0.14^{+0.04}_{-0.03}$ is unlike more massive brown dwarfs per \citet{Bowler20}, but is compatible with formation by both core accretion and disk instability following \citet{Goda19}. Thus the observed parameters of HD~221420~b are plausibly compatible with formation via core accretion (like less massive objects universally considered to be planets) and gravitational collapse (like more massive objects uniformly considered to be brown dwarfs), and whether or not it should be considered a high-mass ``super-planet" or a low-mass brown dwarf is up for interpretation. We do not attempt to adjudicate on this issue here, and instead simply refer to HD~221420~b as a substellar object. Further observations, better understanding of other objects of similar mass, and a more nuanced definition for planets will be necessary to establish a consensus on how to refer to this and other similar companions.

The orbital period of HD~221420~b is comparable to that of Saturn in the Solar system, and coupled with its modest eccentricity it bears some resemblance to the Solar system giant planets, be it at much greater mass. It is then worth considering the possibility of additional interior companions, analogous to the Solar system terrestrial planets, that could be detected using RVs (for example, see \citealt{Barbato18}). Short-period planet detection is somewhat challenging with the available data as the time sampling is relatively sparse, but we are unable to detect any significant short-period signals based on the RV data down to the level of a few \ms{}. This is sufficient to exclude planets with Neptune-like masses on short-period orbits and planets more massive than Jupiter out to several AU, but not to exclude planets of lower mass. Further RV observations would thus be necessary to better constrain the presence of any low-mass interior planets in the system.

\begin{figure*}[ht!]
\figurenum{10}
\epsscale{1.05}
\plottwo{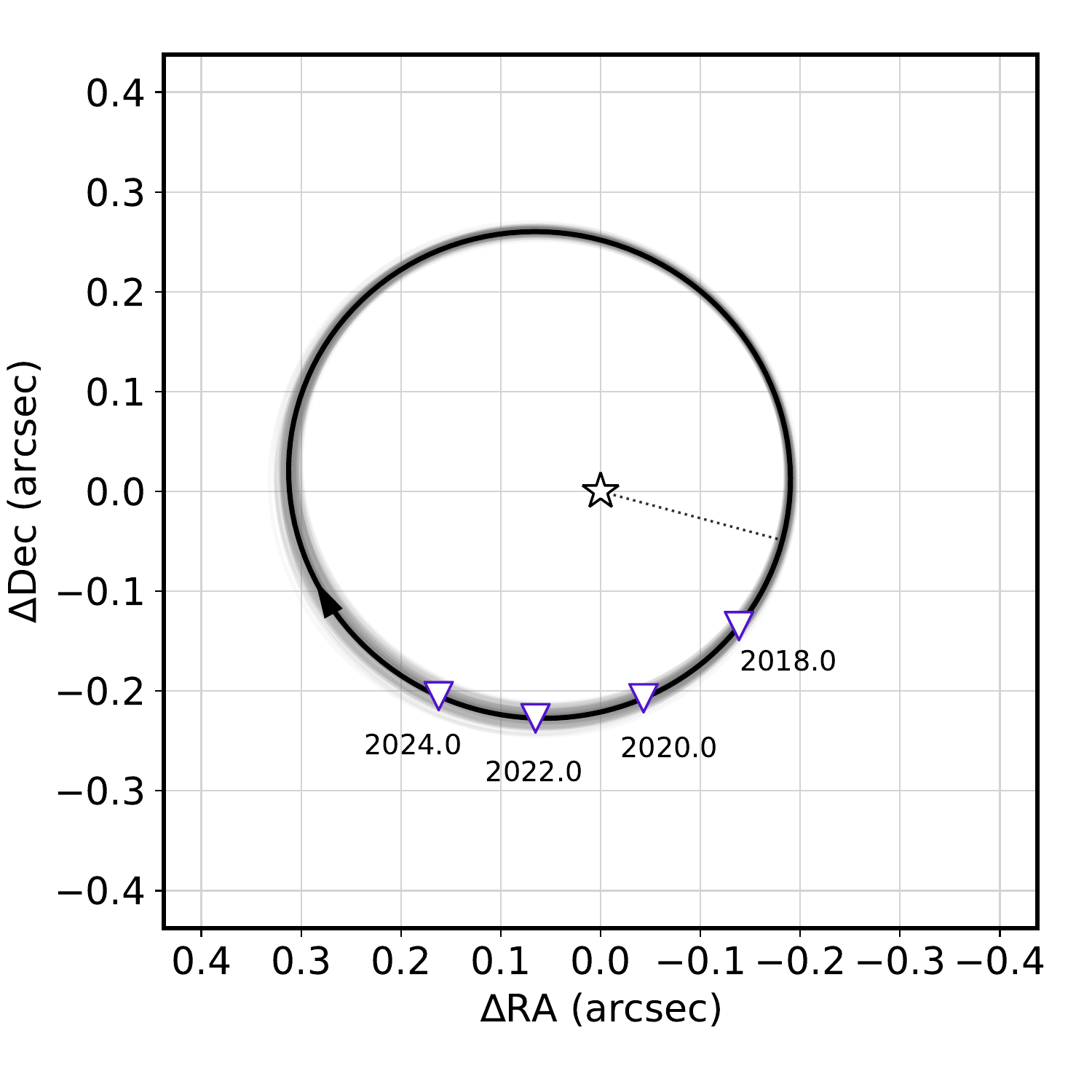}{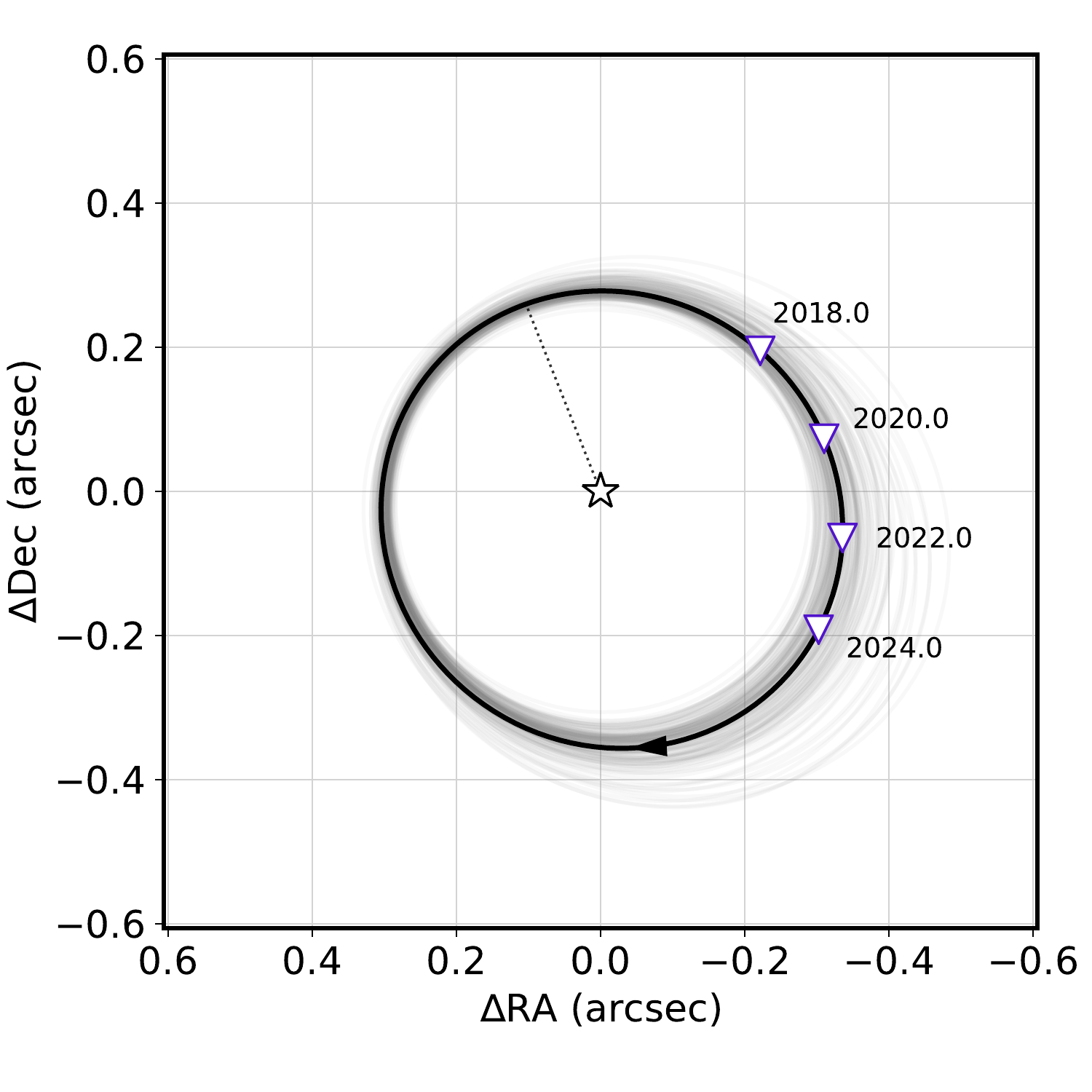}
\caption{Projected relative orbits for HD~92987 (left) and HD~221420 (right) based on the parameters found in this work. The star symbol marks the position of the primary and the dotted line connects it to the periastron of the companion's orbit. The unfilled triangles indicate the predicted positions of the companions for proximate epochs, and the arrow indicates the direction of orbital motion. While the relative orbit of HD~92987~B is tightly constrained, the significant uncertainty in period and eccentricity for HD~221420~b results in large uncertainties on the separation around apoastron. X-axes are reversed to match the direction of positive right ascension on the sky.
\label{figure:both_pos}}
\end{figure*}

\subsection{Prospects for direct imaging}

Both of the companions considered in this work have semi-major axes of around 10 AU, so it is worth considering the potential for their detection using direct imaging. As we have determined the inclinations and longitudes of node for both companions relatively precisely it is possible to project their orbits onto the sky plane, which we show in Figure \ref{figure:both_pos}. More detailed discussion of the prospects for both systems is given below.

Although the stellar nature of HD~92987~B found in this makes it a less interesting target for direct imaging, it is still worth evaluating its potential for detection. Assuming it is a main sequence star HD~92987~B is likely to be a mid-M dwarf, and we expect the contrast with HD~92987~A in the $K_s$ band to be around $\sim$5 magnitudes based on the mass-luminosity relationship of \citet{Mann19}. At a distance of 43.59 $\pm$ 0.07 parsecs the sky semi-major axis of the relative orbit is 250 mas, with a minimum separation of 190 mas and a maximum separation of 320 mas that will next occur around 2032. Although the separation is relatively small, this would be well within the detection capabilities of modern direct imaging instruments.

The parameters of HD~92987~B are already well constrained and imaging of the companion should not be expected to improve on this greatly, but if it is indeed an M-dwarf direct detection would still be useful for placing the companion on the mass-luminosity relationship for low-mass stars (such as \citealt{Mann19}). The 1.2\% precision on the companion mass found in this work is dependent on the model mass of the primary and is therefore unsuitable for the mass-luminosity relation on its own, but sufficient direct imaging observations would make it possible to directly constrain model-independent component masses.

In contrast, direct detection of HD~221420~b is a far more challenging prospect. Although the closer system distance of 31.17 $\pm$ 0.04 parsecs means that the scale of the orbit is larger than HD~92987~B, with a relative semi-major axis of $\sim$330 mas, a minimum separation of $\sim$280 mas, and a maximum separation of $\sim$370 mas which will next occur around 2026 (although the apoastron distance is poorly constrained, and separations as large as 500 mas are possible for long-period solutions), the substellar mass of HD~221420~b means that it should be expected to be faint. Using the \texttt{ATMO} substellar evolutionary models \citep{Phillips20} and assuming an age of \bmaroon{$3.65 \pm 0.32$} Gyr (Table \ref{table:starparams}), we estimate absolute magnitudes for the companion in the 2MASS bands of $(M_J, M_H, M_K) \approx (18.2, 19.2, 19.6)$ mag, resulting in predicted contrasts of $(\Delta_J, \Delta_H, \Delta_K) \approx (15.7, 17.1, 17.8)$ mag. This is near to or beyond the limits of what can be obtained with modern instruments. Taking the example of the SPHERE adaptive optics instrument, contrasts of 16 mag are possible but only at separations of $>$1000 mas \citep{SPHERE}, and contrast limits at the expected separation for HD~221420~b are less stringent; for a practical example, \citet{Maire20.HD72946, Maire20.HD19467} reach $5\sigma$ contrast limits of $\sim5\times10^{-6}$ to $\sim10^{-5}$ ($\approx10-13$ mag) at 350 mas in SPHERE observations of the brown dwarf hosts HD 72946 and HD 19467. The performance of SPHERE on these targets suggests that HD~221420~b would be challenging to detect even at its widest possible separation due to the difficulty of achieving such high contrasts at close separations.

Despite these challenges, direct detection of HD~221420~b would be highly valuable. Its mass is already constrained to 10\% precision - although, as for HD~92987~B, this mass is model-dependent - and combined with the relatively well-determined age of the primary, HD~221420~b would be a benchmark metal-rich substellar object if it can be directly detected. The spectral type of the companion is likely to be approximately late-T, and if it were directly imaged then it would likely be one of the faintest substellar companions to a solar-type star detected so far. Even if HD~221420~b proves undetectable with modern instruments it may become possible to detect with future instrumentation, such as JWST (\citealt{Beichman10}, \citeyear{Beichman19}) or future 30 meter-class telescopes \citep[e.g.][]{Crossfield13, Crossfield16, Artigau18, Quanz19}.

Our estimated absolute magnitudes for HD~221420~b are close to those of Gliese 504 b, an object which was originally identified as $\sim$4 $M_J$ companion to a young star \citep{Kuzuhara13} but was subsequently reinterpreted as a $\sim$23 $M_J$ companion to a peculiarly active turnoff star (\citealt{Fuhrmann15, D'Orazi17, Bonnefoy18}). The revised age and mass of Gliese 504 b are indeed relatively similar to HD~221420~b, although the former lies at a much wider separation from its primary (2.5 arcsec, or $\sim$44 AU projected). If HD~221420~b can be directly detected further comparison between these two objects may prove useful in future.

\section{Conclusions} \label{sec:conclusion}

In this work we have outlined a method for combining radial velocities with the novel astrometric information provided by Hipparcos-Gaia astrometry, and used this to constrain the orbital inclinations of two recently discovered long-period substellar companion candidates.

For HD~92987~B, which has an orbital period of $11640^{+220}_{-190}$ days and a seemingly substellar minimum mass of \bmaroon{$17.08 \pm 0.37$} $M_J$, we measure a near pole-on orbital inclination of $175.82 \pm 0.07 \degree$ and thus a true mass of \bmaroon{$268.4 \pm 4.7$} $M_J$ (\bmaroon{$0.2562 \pm 0.0045$} $M_\odot$), revealing it to be a star masquerading as a substellar object. For HD~221420~b, we find a shorter orbital period of $10090^{+890}_{-560}$ days compared to previous results and an inclination of $164.0^{+1.9}_{-2.6}$ $\degree$, which turns the $6.3^{+0.5}_{-0.3}$ $M_J$ minimum mass into a $22.9 \pm 2.2$ $M_J$ true mass. Owing to \bmaroon{current uncertainty in the distinction} between planets and brown dwarfs, \bmaroon{or indeed if there is any directly observable distinction at all,} HD~221420~b can plausibly be interpreted as a massive planet or a low-mass brown dwarf.

As well as providing measurements of the inclinations, the inclusion of astrometry has led to improvements in the constraints of the physical parameters over those provided by RVs alone for both of the systems studied here. Both of the companions are prospective targets for direct imaging, although we infer that HD~221420~b is most likely too faint to be detectable with modern adaptive optics instruments.

The targets studied here are but two of the over $10^5$ stars included in the Hipparcos-Gaia Catalog of Accelerations (\citealt{Brandt18}, \citeyear{Brandt18erratum}). The principles applied in this work could therefore be extended to provide true masses for many similar systems with long-period companions discovered using RVs, as has been accomplished in a selection of past works. Considering as well the future release of the raw astrometry expected in the Gaia DR4, and it appears that astrometry may soon join the ranks of radial velocities, transit photometry, and direct imaging as fully fledged exoplanet detection techniques.

\acknowledgments

\bmaroon{We thank the anonymous referee for the many helpful suggestions that have improved this manuscript.} Based on observations made with ESO Telescopes at the La Silla Paranal Observatory under programme IDs 072.C-0488, 192.C-0852, 096.C-0499, and 0102.C-0525. This research has made use of the SIMBAD database and VizieR catalogue access tool, operated at CDS, Strasbourg, France. This research has made use of NASA's Astrophysics Data System. This work has made use of data from the European Space Agency (ESA) mission {\it Gaia} (\url{https://www.cosmos.esa.int/gaia}), processed by the {\it Gaia} Data Processing and Analysis Consortium (DPAC, \url{https://www.cosmos.esa.int/web/gaia/dpac/consortium}). Funding for the DPAC has been provided by national institutions, in particular the institutions participating in the {\it Gaia} Multilateral Agreement.

\software{Astropy \citep{Astropy}
	SciPy \citep{SciPy}
	emcee \citep{emcee}
	Matplotlib \citep{Matplotlib}
	corner \citep{corner}
	\bmaroon{kiauhoku \citep{Tayar20}
	lofti\_gaiaDR2 \citep{Pearce20}}
}

\clearpage

\appendix

\section[Appendix A: HD 221420 B]{A widely separated stellar companion candidate to HD 221420} \label{appendix:A}

During the course of this study, we serendipitously discovered a widely separated candidate companion to HD~221420 \bmaroon{that is present in both Gaia DR2 and EDR3}. While the small proper motion of HD~221420 would ordinarily preclude identification of a bound companion, the two stars closely share Gaia proper motions and parallaxes, which significantly increases our confidence in their association. The companion candidate, which we hereon refer to as HD~221420~B, lies at a separation of 698 arcseconds (11.6 arcminutes) from HD~221420~A, translating to a projected separation of 21756 AU (\bmaroon{0.11 pc}). The parallaxes of the two stars agree at the 1$\sigma$ level, \bmaroon{with the DR2 and EDR3 parallaxes differing by $0.10 \pm 0.09$ mas and $-0.035 \pm 0.051$ mas respectively assuming no covariance; the EDR3 values imply a distance difference of only $0.034 \pm 0.049$ parsecs, which when combined with the projected separation implies a physical separation of $<37800$ AU ($<0.18$ pc) at 99\% confidence. This is a relatively wide separation for a binary, but within the range of known systems (\citealt{Shaya11, Mamajek13}).}

We list parameters of the two stars in Table \ref{table:appA1}. HD~221420~B does not have an established spectral type, but we infer it is likely to be a mid- or late-M dwarf. \bmaroon{Recently \citet{Sebastian21} have produced photometric parameters for cool stars and brown dwarfs within 40 parsecs; for HD~221420~B they report an estimated spectral type of M$4.6\pm0.8$ and an effective temperature of $3065 \pm 105$ K. Spectroscopic confirmation of these parameters remains desirable, but a spectral type around $\sim$M5V appears to be plausible.}

A cause for concern is that HD~221420~B lies near to a brighter star, Gaia DR2 6377398274119547520, a background G-dwarf \bmaroon{2 magnitudes brighter in the G-band that lies at a separation of only 3.7 arcsec}. This is close enough that light contamination could conceivably be an issue, and there is some potential evidence for this in that HD~221420~B has a statistically significant astrometric excess noise of 0.62 mas in Gaia DR2 \bmaroon{and 0.26 mas in Gaia EDR3}. However, the Renormalised Unit Weight Error (RUWE) for the \bmaroon{two} astrometric solutions are only 1.25 \bmaroon{and 1.16 respectively}, which \bmaroon{are} not greatly higher than the expected value of 1.0 and is below the proposed upper limit for a ``well-behaved" source of 1.4 \citep{GaiaDR2}. Thus, although there could be some mild contamination from the brighter neighbouring star, the reported parallax and proper motions for HD~221420~B are likely to be adequately accurate.

\bmaroon{To quantify the probability of chance alignment, we query Gaia EDR3 for all sources within a 30 degree radius of HD 221420 A that have a parallax within $\pm$ 1 mas of HD 221420 A's parallax. This search returned 65 sources. The probability of finding two stars within this range of parallaxes within 700 arcseconds of eachother is therefore only $0.0014$. We further refine this search by querying the same area for stars that have proper motions within $\pm$ 5 \masyr{} of HD 221420 A's proper motion in both axes. This search returns only HD~221420~A and B, indicating that the close alignment in position, parallax, and proper motion between these two stars is vanishingly unlikely to arise by chance.}

\bmaroon{Direct comparison of the proper motions of HD~221420~A and B is complicated by the fact that their considerable separation makes projection effects non-negligible. \citet{ElBadry19} investigated the significance of projection effects on the relative velocities of wide binaries and found that for binaries within 120 pc the discrepancy between the apparent and true velocity differences become significant beyond $\gtrsim0.1$ parsecs, a value which is comparable to the observed separation between HD~221420~A and B; it is therefore important to attempt to correct for projection effects before interpreting the relative velocity of the two stars.}

\bmaroon{HD~221420~B does not presently have a radial velocity measurement, precluding a direct comparison of the space velocities of the two stars. To correct for projection effects in such systems \citet{ElBadry19} assumed equal velocities for both components, but this introduces an assumption of physical association that is not warranted in this case. Instead, we take a different approach to ``de-project" the relative proper motion difference, using a technique recently used by \citet{Tofflemire21} to identify members of a moving group. As the barycentric radial velocity and proper motions of HD~221420~A are known precisely (Table \ref{table:HD221420}), it is straightforward to calculate its space velocities using the equations given in \citet{Johnson87}. These can then be used to calculate the proper motion of HD~221420~A that would result if the star instead lay at \textit{the position of HD~221420~B}; this provides accurate proper motions for A normalised to the reference frame of B, thus correcting for projection effects, which can then be compared to the observed proper motion of B in the usual manner. This therefore offers a way estimate the true tangential velocity difference between the two stars without requiring a radial velocity measurement for HD~221420~B.}

\bmaroon{Applying this to HD~221420~A, we find an expected proper motion if the star was at the position of B of ($15.87 \pm 0.16$, $0.07 \pm 0.15$) \masyr{}, a significant difference of ($0.87$, $-0.58$) \masyr{} from the directly observed barycentric proper motion. Comparing this ``de-projected" proper motion with the proper motions of HD~221420~B in Gaia DR2 and EDR3, we find (B$-$A) proper motion differences in RA and declination of ($-1.63 \pm 0.31$, $-1.89 \pm 0.27$) and ($-0.94 \pm 0.21$, $-1.42 \pm 0.20$) \masyr{} respectively, equivalent to projected tangential velocity differences of $370 \pm 60$ and $250 \pm 30$ \ms{}. We estimate the escape velocity of the system to be $\lesssim350$ \ms{} so these tangential velocities are plausibly compatible with bound orbits, although they imply relatively high orbital velocities.}

\begin{deluxetable}{ccc}[t!]
\centering
\tablecaption{Parameters of HD 221420 and its candidate companion.\label{table:appA1}}
\tablehead{\colhead{Parameter} & \colhead{HD 221420 A} & \colhead{HD 221420 B}}
\startdata
Gaia DR2\bmaroon{/EDR3} ID & 6353376831270492800 & 6377398274119547392 \\
2MASS ID & J23331963-7723069 & J23315632-7712250 \\
\hline
Gaia G-band magnitude [mag] & \bmaroon{5.656} & \bmaroon{15.606} \\ 
Gaia BP-RP colour [mag] & \bmaroon{0.831} & \bmaroon{3.494} \\
Renormalised Unit Weight Error (RUWE) & \bmaroon{1.15} & \bmaroon{1.16} \\
Parallax $\varpi$ [mas] & \bmaroon{$32.102 \pm 0.033$} & \bmaroon{$32.067 \pm 0.039$} \\
RA proper motion $\mu_{RA}$ [\masyr{}] & $+15.00 \pm 0.06$ & \bmaroon{$+14.93 \pm 0.05$} \\
Dec proper motion $\mu_{Dec}$ [\masyr{}] & $+0.65^{+0.08}_{-0.05}$ & \bmaroon{$-1.35 \pm 0.05$} \\
\hline
\bmaroon{Mass [${M_\odot}$]} & \bmaroon{$(1.351 \pm 0.017) + (0.022 \pm 0.002)$} & \bmaroon{$0.134 \pm 0.006$} \\
\hline
Projected separation [arcsec] & -- & \bmaroon{$698.42349 \pm 0.00004$} \\
Position angle [degrees] & -- & \bmaroon{$336.843233 \pm 0.000001$} \\
Projected separation [AU] $^{(a)}$ & -- & \bmaroon{$21756 \pm 22$} \\
Projected tangential velocity difference [\ms{}] $^{(a)}$ & -- & \bmaroon{$250 \pm 30$} \\
\enddata
\tablerefs{All data is derived from Gaia EDR3 \citep{GaiaEDR3} except the proper motion parameters of HD~221420~A which are from Table \ref{table:HD221420} and the masses, which are discussed further in the text.}
\tablecomments{(a) Assuming the system lies at the distance implied by the primary's parallax; these values are thus lower limits for the true values.}
\end{deluxetable}

To investigate the possibility of constraining the orbit of the binary we use \texttt{lofti\_gaiaDR2} \citep{Pearce20}, \bmaroon{a code designed to constrain the orbital parameters of wide binaries using Gaia astrometry}. For the fit we use the astrometric parameters \bmaroon{from Gaia EDR3 as} listed in Table \ref{table:appA1}, \bmaroon{except for the proper motion of A where we use the ``de-projected" values given above}. \bmaroon{We adopt a mass} for HD~221420~A of \bmaroon{$1.373 \pm 0.019$} $M_\odot$ (the sum of the masses of the primary star and its substellar companion, disregarding the negligible covariance between the two values), and a mass of $0.134 \pm 0.006$ $M_\odot$ for HD~221420~B based on the mass-luminosity relationship of \citet{Mann19}. \bmaroon{We are, however, unable to find adequate orbital solutions for the provided data, as the best-fit orbits found by \texttt{lofti\_gaiaDR2} have unacceptably high $\chi^2$ values. Examination of the posteriors shows that the best fit orbits are skewed towards very high eccentricities ($\gtrsim$0.9) and place HD~221420~B close to periastron on orbits with unreasonably large semimajor axes ($\sim$1 pc). Thus, while the observed tangential velocity difference between HD~221420~A and B is consistent with bound orbits, the velocity difference appears to be too high to produce ones which are physically plausible.}

\bmaroon{A possible explanation for this discrepancy is that HD~221420~B has an undetected close companion. As the star has no history of observation this is entirely possible, and a companion (stellar or substellar) that produces a tangential velocity anomaly of only a few hundred \ms{} would be sufficient to produce reasonable orbits for HD~221420~AB.} Indeed, were it not for the Hipparcos and RV observations of HD~221420~A there would be no reason to suspect the presence of a substellar companion based on Gaia DR2 alone, and we would not be able to extract the ``true" barycentric proper motion of the star without more data. \bmaroon{\citet{Clarke20} considered the impact of unresolved triples on the apparent tangential velocities of wide binaries in Gaia DR2 and found that this is sufficient to explain the observed population of seemingly unbound binaries. We therefore consider it plausible that the peculiarly large tangential velocity difference between HD~221420~A and B could be caused by an unresolved companion of the latter star. Potential supporting evidence for this hypothesis is non-zero astrometric excess noise for HD~221420~B as discussed above as well as its significant change in proper motion between the Gaia DR2 and EDR3 catalogues, with the two solutions differing in RA and declination by ($+0.69, +0.47$) \masyr{}, several times the stated uncertainties. Future Gaia releases may reveal the presence of an astrometric companion to HD~221420~B, which could resolve the high tangential velocity anomaly presently observed.}

In conclusion, we have discovered a candidate red dwarf companion to HD~221420. \bmaroon{The two stars are physically separated by only $<0.18$ parsecs and are consistent with a bound pair, but the observed difference in proper motions is difficult to reconcile with plausible orbits; we suggest that this discrepancy could be resolved by the presence of an unresolved companion to the secondary star, which may be detectable by Gaia. Further observations of both stars will be necessary to confirm or reject the nature of their association, but we treat the pair as a candidate binary here.}

\bmaroon{Subsequent to the time of writing the candidate companion discussed here was independently detected by \citet{ElBadry21}, who constructed a large catalogue of resolved binaries in Gaia EDR3. In that work the authors estimate the chance alignment probability for a given binary by comparing the detected systems to an artificial catalogue, and for HD~221420~AB this procedure results in a low chance alignment probability of $1.6\%$. This independent detection increases our confidence that the two stars form a genuine wide binary.}

\section[Appendix B: HARPS RVs]{HARPS radial velocities} \label{appendix:B}

Here we tabulate the HARPS data used in this work. As discussed in Section \ref{subsec:data} the HARPS RVs taken before the 2015 instrument upgrade (``HARPS-03") have been extracted from \citet{Trifonov20}, while the measurements taken subsequent to the 2015 HARPS upgrade (``HARPS-15") have been downloaded from the ESO science archive. As HARPS RVs are natively barycentric but \citet{Trifonov20} subtracted a constant offset from their re-reduced RV data, the RVs in Table \ref{table:HARPSpre} are relative while the RVs in Table \ref{table:HARPSpost} are absolute. Finally, we have taken nightly bins of the RVs to reduce the amount of epochs for the model; this is justifiable here as we are not concerned with short period ($<$1 day) variability in this study.

\clearpage

\begin{deluxetable}{ccc}[h!]
\centering
\tablecaption{HARPS-03 RVs used in this work.\label{table:HARPSpre}}
\tablehead{\colhead{Time [BJD]} & \colhead{RV [\ms{}]} & \colhead{Error [\ms{}]}}
\startdata
2452937.60122 & 13.70  & 0.99 \\
2453146.93312 & 8.46   & 0.99 \\
2453149.93522 & 15.64  & 1.09 \\
2453151.95139 & 10.91  & 0.84 \\
2453202.83636 & 11.34  & 1.00 \\
2453265.73680 & 9.25   & 0.94 \\
2453266.68728 & 9.74   & 1.01 \\
2453267.66919 & 9.84   & 0.98 \\
2453269.72460 & 10.95  & 1.13 \\
2453295.69296 & 4.50   & 0.95 \\
2453339.56872 & 9.93   & 0.92 \\
2453543.90019 & 2.92   & 0.71 \\
2453544.92535 & 2.53   & 0.93 \\
2453547.92975 & 2.09   & 0.94 \\
2453550.92202 & 3.24   & 0.99 \\
2453551.92843 & 3.86   & 0.95 \\
2453574.82083 & -0.35  & 0.81 \\
2453575.81902 & 1.24   & 0.63 \\
2453576.85061 & 5.32   & 0.95 \\
2453579.80920 & 3.35   & 0.93 \\
2453869.92198 & -5.48  & 0.93 \\
2454314.78316 & -18.13 & 1.03 \\
2454316.82337 & -18.58 & 0.80 \\
2454319.84250 & -20.00 & 0.96 \\
2456621.51422 & -42.45 & 0.75
\enddata
\end{deluxetable}

\begin{deluxetable}{ccc}[h!]
\centering
\tablecaption{HARPS-15 RVs used in this work.\label{table:HARPSpost}}
\tablehead{\colhead{Time [BJD]} & \colhead{RV [\ms{}]} & \colhead{Error [\ms{}]}}
\startdata
2457336.68701 & 26568.84 & 0.24 \\
2457338.73313 & 26572.19 & 0.39 \\
2457339.66100 & 26568.71 & 0.29 \\
2457340.63632 & 26569.96 & 0.21 \\
2457360.54115 & 26568.57 & 0.19 \\
2457361.54315 & 26568.12 & 0.19 \\
2457362.54317 & 26570.09 & 0.14 \\
2458419.76134 & 26607.58 & 0.49 \\
2458429.53695 & 26605.57 & 0.23 \\
2458430.58013 & 26610.90 & 0.23
\enddata
\end{deluxetable}

\clearpage

\bibliography{bib}
\bibliographystyle{aasjournal}


\end{document}